# Passive Cavitation Mitigation on Hydrofoils via Porous Media: A Comparative Study of LES and RANS Models

Ali Alavi [a], Maziyar Ghasemnezhad [a], Ali Sangtarash [a], Ehsan Roohi [b*]

[a] High Performance Computing (HPC) Laboratory, Department of Mechanical Engineering, Faculty of Engineering, Ferdowsi University of Mashhad, P.O. Box 91775-1111, Mashhad, Iran

[b] Mechanical and Industrial Engineering, University of Massachusetts Amherst, 160 Governors Dr., Amherst, MA 01003, USA. (corresponding author: roohie@umass.edu)

**Abstract**

This study numerically investigates the use of porous media as a passive strategy for mitigating cavitation on a NACA 66 (MOD) hydrofoil subjected to unsteady two- phase flow. Employing the Volume of Fluid (VOF) method alongside the Schnerr–Sauer cavitation model, simulations are performed at two cavitation numbers ($\sigma = 1$ and $0.7$) for both 2D and 3D geometries. The porous region, characterized by a porosity of 0.95. and spanning one-third of the suction side, is represented as a momentum sink using Ergun's equation in ANSYS Fluent. Three turbulence models are utilized in the 2 D simulations: Large Eddy Simulation (LES), realizable k–$\varepsilon$, and k–$\omega$ SST turbulence models to evaluate their predictive performance. For the 3 D simulations, the LES turbulence model is employed to capture three- dimensional flow and validate the effectiveness of the porous media. The treatment of the porous surface influences cavitation dynamics by stabilizing the boundary layer, minimizing sharp pressure gradients, and limiting the growth of cavitation bubbles. Contour visualizations of the pressure coefficient, velocity magnitude, and vapor volume fraction indicate that the porous layer reduces negative pressure near the wall, lowering local suction intensity and delaying cavitation inception. As a result, the formation of re-entrant jets and aggressive cavity shedding- dominant in non-porous setups- is mitigated. Additionally, distributions of turbulent kinetic energy (TKE), z- vorticity, and baroclinic torque illustrate that the porous media diminishes vortex strength and wake instabilities. Spectral analyses of drag and lift coefficients through Fast Fourier Transform (FFT) reveal a notable reduction in high- frequency components for the porous case, demonstrating its ability to mitigate unsteady hydrodynamic loads. Collectively, these findings highlight the potential of porous integration as a robust and energy- efficient approach to enhance hydrofoil stability and cavitation resistance in marine systems.

**Nomenclature**

*Abbreviations*

| | | | |
|---|---|---|---|
| LES | Large Eddy Simulation | $K$ | turbulent Kinetic Energy |
| VOF | Volume of Fluid | $\bar{S}$ | Modulus of the Rate of Strain |
| SGS | subgrid-Scale | *Greek letters* | |
| GS | grid-Scale | $\sigma$ | cavitation number |
| RANS | reynolds-Averaged Navier-Stokes | $\rho$ | density |
| SST | Shear Stress Transport | $\nu$ | kinematic viscosity |
| FFT | Fast Fourier Transform | $\mu$ | dynamic viscosity |
| AOA | angle of Attack | $\alpha$ | angle of Attack |
| TKE | turbulent Kinetic Energy | $\nu_t$ | subgrid-Scale Eddy viscosity |
| $n_0$ | Initial Number of Bubbles Per Unit Volume | $\tau_{ij}$ | Subgrid-Scale Stresses Tensor |
| $P_g$ | vapor Pressure | $\gamma$ | vapor Fraction |
| $P_\infty$ | ambient Pressure | $\varepsilon$ | turbulence Dissipation Rate (k-ε model) |
| $P_l$ | liquid Pressure | $\omega$ | Specific Dissipation Rate (k-ω model) |
| $V_\infty$ | flow Speed | $\sigma_\omega, \sigma_k, \sigma_\varepsilon$ | diffusion Coefficients |
| $\tilde{D}$ | Shear Stress Tensor | $\Gamma$ | Auxiliary Variable in Blending Function |
| $\dot{m}$ | Mass Flow Rate | $\kappa$ | von Kármán Constant |
| $R_b$ | bubble Radius | $\phi$ | porosity |

| | |
|---|---|
| $\bar{g}$ | Gravitational acceleration |
| $C_\mu$ | Turbulent Viscosity Coefficient (k-ε model) |
| $P_k$ | turbulent Kinetic Energy Production Term |
| $CD_{k\omega}$ | Cross-Diffusion Term in k-ω SST |

## 1. Introduction

Cavitation refers to the development and sudden collapse of vapor bubbles within a liquid, occurring when local pressure falls below vapor pressure. This phenomenon is often represented by the Rayleigh–Plesset equation. It's frequently observed in the vicinity of marine propellers, pumps, and hydrofoils, where it generates intense microjets and shock waves that can damage surfaces. There are various forms of cavitation, including sheet, cloud, vortex, and traveling bubble cavitation, each type inducing distinct flow disruptions and structural issues. If not adequately managed, cavitation can lead to severe erosion, elevated noise levels, vibrations from flow, and unstable hydrodynamic loads, which can ultimately undermine system performance and structural integrity.

In recent years, the numerical simulation of cavitating flows has emerged as a significant area of research, focusing on capturing the unsteady and nonlinear dynamics of cavity evolution. Ji et al. [10, 11] showed that strong fluctuations in cavity volume are mainly driven by interactions between vortices and cavities, as well as re-entrant jet behavior. Aktaş et al. [12] connected cavitation regimes to propeller noise, highlighting the role of simulation in noise control. Li et al. [13] confirmed the reliability of simulations through experimental validations. Guo et al. [14] concentrated on tip leakage vortex cavitation to inform propeller design. Kolahan et al. [15] examined the propagation of cavitation waves around a sphere, concluding that the intensity of flow oscillations rose with an increasing number of cavitation events. Zahiri and Roohi [16] utilized the AMD sub-grid model in OpenFOAM, enhancing predictions of friction, velocity, and turbulence. Wang et al. [17] further emphasized that unsteady cavitation could generate compression-induced shock waves, highlighting the necessity for high-resolution models to achieve accurate predictions. Chen et al. [18] introduced linear vortex models to improve numerical efficiency, while Rashidi et al. [19] implemented finite element methods to refine cavitator design and enhance cavitation control.

Hydrofoils, utilized in high-speed marine transport and renewable energy systems, create lift while minimizing drag. However, they are highly susceptible to cavitation. Vapor cavities often form on the suction side of the hydrofoil, disrupting the lift distribution, inducing unsteady forces, and accelerating surface erosion. This ultimately reduces efficiency and shortens service life [6, 20]. Cavitation-related issues are particularly significant in high-performance or stealth-sensitive applications where maintaining surface integrity and flow stability is crucial. Therefore, effective cavitation control is essential to preserve durability and performance in these scenarios. Guo et al. [21] observed strongly unsteady, periodic cavitation patterns in viscous flow conditions around hydrofoils, while Ji et al. [22] demonstrated a direct correlation between pressure fluctuations and the cavity shedding process on a NACA66 hydrofoil. These studies underscore the importance of controlling cavitation to prevent erosion, noise, flow-induced vibrations, unsteady loads, and shock waves. Such disruptions can negatively impact lift, damage surfaces, and ultimately degrade overall system performance and longevity.

To reduce cavitation, researchers have investigated both passive and active control methods. Passive control employs fixed geometric or surface modifications that alter the flow field and minimize cavitation without needing external energy input [23]. Examples include bio-inspired riblets, surface roughness at the tips, and barriers on the suction side, which disrupt the flow instabilities that cause the formation and collapse of cavities [24, 25]. Movahedian et al. [26] discovered that a lower cavitation number leads to an increase in the size of trailing cavities in twisted hydrofoils. Pendar et al. [27] demonstrated that a sinusoidal leading edge (SLE) on a NACA 634-021 hydrofoil can delay bubble separation. These solutions are straightforward, reliable, and require minimal maintenance.

Active control methods, in contrast, utilize external energy or real-time actuation to manipulate vapor structures during operation. A widely studied active technique is forced ventilation, where gas is injected into the flow to replace vapor cavities and stabilize the fluid around the hydrofoil [28]. While practical, active approaches usually exhibit mechanical complexity and high energy demands, which may render them less suitable for certain applications. On the other hand, several passive techniques, including tubercles, riblets, and vortex generators, have been investigated to mitigate cavitation [29-31]. However, many of these solutions struggle to achieve an effective balance between cavitation mitigation and aerodynamic performance. One notable passive control method is porous media, which incorporates permeable structures into surfaces to influence local pressure distributions and suppress cavitation dynamics without needing external input.

Porous media consist of heterogeneous materials containing interconnected voids that enable the transport of fluids, heat, and mass. These internal voids facilitate pressure-driven flow, which is essential for numerous natural and engineered systems. The functionality of these materials is dictated by porosity, quantifying the proportion of void space and reflecting the storage capacity, as well as permeability, which gauges the ease of fluid movement under a pressure gradient [32, 33]. Natural porous substances like soil and rock are vital in groundwater management and resource extraction [34], while engineered variants, such as foams and scaffolds, play roles in filtration, heat exchange, and biomedical applications [35]. In addition to fluid transport, porous media improve heat transfer and support catalytic processes and tissue growth due to their high surface-to-volume ratio [36-39]. Recent developments in cavitation control have highlighted the effectiveness of porous materials as passive methods for suppressing cavity growth, alleviating vortex shedding, and stabilizing unsteady pressure fields in various flow conditions. For instance, Bi et al. [40] illustrated that porous tip treatments in hydrofoils diminish the strength of the tip leakage vortex and stabilize the flow, with performance affected by the size of the gap. Yu et al. [41] reported enhancements in pressure near the leading edge and a delay in cavity formation when using porous layers on a hemispherical cylinder, while Sadri et al. [42] found that coatings with adjusted permeability on a circular cylinder minimize lift fluctuations and frequency of cavitation shedding. Qiu et al. [43] reviewed both passive and active techniques, underscoring the energy efficiency of porous treatments. Yu et al. [44] noted that suppression performance achieves its maximum at moderate cavitation numbers but decreases significantly at very low σ. Overall, properly configured porous media effectively serve as passive cavitation control mechanisms.

The cavitation number (σ) is a key dimensionless parameter used to evaluate the likelihood of cavitation, with higher values indicating lower susceptibility, calculated by $\sigma = P_\infty - P_g / 0.5 \rho V_\infty^2$ [45, 46]. Accurate prediction of cavitating flows requires advanced mass transfer models, interface-capturing schemes, and appropriate turbulence models. Sauer has refined classical Rayleigh-based models [47] and Yuan et al. [48], while semi-empirical approaches by Merkle et al. [49], Kunz et al. [50], and Singhal et al. [51] offer alternative formulations. Senocak et al. [52] introduced a model based on mass and momentum conservation across the cavity interface.

Among the modeling techniques, the Volume of Fluid (VOF) method stands out for its effectiveness in simulating cavitating flows. It accurately tracks the liquid-vapor interface as a sharp discontinuity through the use of advection equations, allowing for precise reconstruction of cavity shapes. VOF has demonstrated versatility in various multiphase flow situations, including droplet impacts, spray dynamics, and liquid sloshing [53, 54]. In particular, studies on cavitation by Ait Bouziad et al. [55] and Wiesche [56] confirmed its effectiveness for flows involving hydrofoils and impellers, highlighting its ability to accurately capture the formation and structure of vapor cavities.

In cavitating flow simulations, the choice of turbulence model critically impacts predictive accuracy. This study compares three models: This study compares three models: Large eddy simulations (LES), K-epsilon (k-ε), and Shear Stress Transport k-ω (SST k-ω), balancing fidelity and computational efficiency [57]. LES explicitly resolves the large-scale, energy-containing eddies directly interacting with vapor structures, making it highly effective for capturing key unsteady features such as cavity shedding, re-entrant jet formation, and shock waves produced during cavity collapse [22, 58]. LES is often employed in conjunction with mass transfer models like Zwart–Gerber–Belamri (ZGB) or Schener–Sauer to simulate vapor–liquid interaction dynamics accurately [59, 60]. For instance, Long et al. [61] used LES with a homogeneous cavitation model to simulate turbulent cavitating flows around a sphere, successfully resolving intricate vortex structures and cavity shedding behaviors.

Conversely, the k-ε model addresses the transport equations for turbulent kinetic energy and its dissipation rate. This full-spectrum turbulence model is computationally efficient, albeit less detailed. It excels in predicting mean flow properties such as cavity geometry, lift and drag coefficients, and shedding frequencies, making it a valuable option for numerous engineering applications [62, 63].

The SST k-ω model employs a hybrid approach that merges the near-wall resolution of the k-ω formulation with the strong free-stream characteristics of the k-ε model. This combination improves performance in areas with adverse pressure gradients commonly found in cavitating flow environments, leading to more dependable and precise simulations [57].

This study presents an innovative passive strategy for cavitation control by incorporating porous media into the hydrofoil surface. Porous materials serve as passive flow modifiers, facilitating partial pressure equalization, altering boundary layer behavior, and dissipating the energy of re-entrant flow that triggers cavity shedding. These effects can postpone cavitation onset, enhance cavity dynamics stability, and mitigate aggressive bubble collapse without the need for external energy. To assess the efficacy of this method, we utilize three widely recognized turbulence models—LES, k-ε, and SST k-ω—each providing unique benefits in capturing flow unsteadiness, turbulence dynamics, and cavity development. The outcomes of this research could revolutionize hydrofoil design by establishing a robust and energy-efficient approach for enhancing performance and durability in cavitating flow conditions.

The rest of this paper is structured as follows: Section 2 discusses the governing equations and physical models, highlighting the VOF method for interface tracking and the application of three turbulence models: LES, k-ε, and SST k-ω. Additionally, it describes the approach for integrating and modeling porous media on the hydrofoil surface. Section 3 focuses on the numerical setup, boundary conditions, mesh sensitivity, and validation methods. In Section 4, we present results for both porous and non-porous hydrofoils across different cavitation numbers, examining pressure distribution, fluctuations in lift and drag forces, spectral analysis using Fast Fourier Transform (FFT), baroclinic torque, vorticity, and turbulence characteristics. We evaluate the effects of turbulence modeling and the design of porous surfaces, including the role of porous media in suppressing cavity shedding, dampening vortex structures, and enhancing the overall performance of hydrofoils under unsteady cavitating conditions. Section 5 concludes with a summary of findings and proposes future research directions in passive cavitation control using porous materials.

## 2. Governing equations

The interface between liquid and vapor phases, the cavitation boundary, defines the free surface in cavitating flows. To describe the fluid behavior across this interface, linear conservation laws are formulated for a control volume moving with an arbitrary velocity, based on the principles of displacement theory. This formulation leads to the governing equations of continuity and momentum (Navier–Stokes) as expressed in equations (1) and (2) [51].

$$\vec{\nabla}\vec{u} = \dot{m}(\frac{1}{\rho_l} - \frac{1}{\rho_g}) \tag{1}$$

$$\frac{\partial}{\partial t}(\rho\vec{u}) + \vec{\nabla}.(\rho\vec{u}\vec{u}) = -\vec{\nabla}p + 2\vec{\nabla}.\left[\mu\vec{D}\right] + \rho\vec{g} + \sigma\kappa\vec{\nabla}\gamma_1 \tag{2}$$

In these equations, $\vec{u}$ (m/s) denotes velocity, $P$ (Pa) represents pressure, and $\vec{g}$ (m/s$^2$) signifies gravitational acceleration. The shear stress tensor magnitude is defined according to equation (3) [15]. These governing equations assume that the fluid phases are incompressible and that both liquid and vapor phases share the same initial velocity.

$$\vec{D} = \frac{1}{2}\left[\vec{\nabla}\vec{u} + \left(\vec{\nabla}\vec{u}\right)^T\right] \tag{3}$$

The governing equations are derived assuming that both liquid and vapor phases are incompressible and initially possess the same velocity. Within this framework, the symbol $\vec{u}$ (m/s) denotes the magnitude of the time-averaged mixture velocity, while $P$ (Pa) represents the time-averaged pressure field. $\dot{m}$ (kg/s) denotes the mass transfer rate associated with phase change during cavitation events. Turbulence effects are modeled using a viscosity coefficient $\mu_t$ (Pas), which accounts for eddy-induced momentum transport. The volume fractions of the liquid ($\gamma_t$ (no unit)) and vapor ($\gamma_g$ (no unit)) phases are also included in the formulation and are defined through the following equations.

$$\gamma_t = \frac{\dot{m}_t}{\dot{m}_{Total}} \tag{4}$$

$$\gamma_g = \frac{\dot{m}_g}{\dot{m}_{Total}} \tag{5}$$

Equations (5) and (6) define the total mass flow rate, denoted by $\dot{m}_{Total}$ (kg/s), as shown in equation (6).

$$\dot{m}_{Total} = \dot{m}_t + \dot{m}_g \tag{6}$$

Consequently, the mixture viscosity ($\mu$ (Pas)) and density ($\rho$ (kg/m³)) are defined based on this term as follows [64]:

$$\begin{aligned}\rho &= \gamma_t \rho_t + (1-\gamma_t)\rho_g \\ \mu &= \gamma_t \mu_t + (1-\gamma_t)\mu_g\end{aligned} \tag{7}$$

## 2.1. Mass transfer model

In cavitating flows, the interphase mass transfer is dictated by a transport equation. This equation incorporates the fluid volume fraction and a mass source term, which enables the exchange between liquid and vapor phases, as shown in the equation (8).

$$\frac{\partial \gamma_l}{\partial t} + \nabla.(\gamma_l u) = (1-\gamma)\frac{4\pi R_b^2 n_0}{1+\frac{4}{3}\pi R_b^3 n_0}\frac{DR_b}{Dt} \tag{8}$$

$n_0$ denotes this equation's initial number of vapor nuclei per unit volume. The present study employs the Sauer model [47, 65], which expresses the vapor fraction ($\gamma$) as a function of the number of vapor nuclei per unit volume and their diameter ($R_b$), which is assumed to be uniform. This approach extends classical cavitation theory by recognizing that the vapor content depends on bubble size and the quantity of nuclei in the flow field [48]. Based on these assumptions, the mass transfer rate is expressed through equation (9).

$$\gamma = \frac{V_g}{V_g + V_l} = \frac{\frac{4}{3}\pi R_b^3 n_0}{1 + \frac{4}{3}\pi R_b^3 n_0} \tag{9}$$

By simplifying the model and neglecting second-order derivative terms, the governing relations reduce to equations (10) and (11), where $P_g$ (Pa) is the vapor pressure and $\bar{P}$ (Pa) represents the filtered (or resolved) pressure field within the liquid.

$$\frac{DR_b}{Dt} = -sign(\bar{P} - P_g)\sqrt{\frac{2}{3}\frac{|\bar{P} - P_g|}{\rho_l}} \tag{10}$$

$$\dot{m} = -\rho_g(1-\gamma)\frac{3\gamma}{R_b}sign(\bar{P} - P_g)\sqrt{\frac{2}{3}\frac{|\bar{P} - P_g|}{\rho_l}} \tag{11}$$

To perform the simulation, initial values for the nuclei number density and diameter ($d_{Nuc}$) are specified as $n_0 = 10^{13}$ (nuclei/m$^{-3}$) and $d_{Nuc} = 2\times 10^{-6}$ (m), respectively. These default parameters are recommended by Ansys Fluent and widely accepted for their reliability in cavitation modeling [66]. Under these assumptions, the $R_b$ can be expressed as a function of these parameters using equation (12).

$$R_b = \left( \frac{1}{\frac{4}{3}\pi n_0} \frac{\gamma}{1-\gamma} \right)^{1/3} \tag{12}$$

## 2.2. Volume of fluid (VOF) method

The VOF method reconstructs the interface between liquid and vapor phases in cavitating flows. This method introduces a scalar field $\gamma$ that represents the volume fraction of fluid within each computational cell and is presented in equation (13) [67].

$$\gamma = \begin{cases} 1 & \text{fluid 1(Liquid)} \\ 0 & \text{fluid 2(Vapor)} \\ 0 < \gamma < 1 & \text{at the interface} \end{cases} \tag{13}$$

To accurately preserve the sharpness of the interface, a compressive velocity term $u_c$ is incorporated into the VOF formulation, as shown in equation (14). The normalized mass flow rate ($\bar{m}$) in this equation is determined by the underlying mass transfer model, typically based on the methodologies introduced by Kunz and Sauer. The compressive velocity component $u_c$, initially proposed by Rusche [68], acts exclusively near the interface and functions as a localized surface compression term.

$$\frac{\partial \gamma}{\partial t} + \nabla \cdot (\gamma \vec{u}) + \nabla \cdot [\vec{u}_c \gamma (1-\gamma)] = \bar{m} \tag{14}$$

VOF methods generally integrate first-order limited downwind schemes with higher-order accurate approaches to minimize numerical diffusion while preserving resolution. This technique is particularly useful in predicting impact pressures from collapsing bubbles and incorporates surface tension models that aid in defining bubble shapes. Ye et al. [69] and Zhang et al. [70]

validated its effectiveness in capturing cavitation dynamics. More details on compressive VOF formulations can be accessed in [67, 71].

**2.3. Large eddy simulation (LES)**

Large Eddy Simulation (LES) models turbulence by directly resolving significant large-scale structures that carry most of the flow's energy, while addressing the smaller eddies with subgrid-scale (SGS) modeling [72]. By applying a low-pass filter to the Navier–Stokes equations, LES distinguishes itself from RANS by effectively capturing a wide range of flow scales. This capability makes LES particularly useful for simulating unsteady cavitating flows [73, 74]. The governing equations of continuity and momentum for LES in the context of incompressible, multiphase, and unsteady flows are defined by Equations (15) and (16). In these equations, $p$ represents the mixture pressure, while $u_i$ denotes the velocity field.

$$\frac{\partial \rho_m}{\partial t} + \frac{\partial (\rho_m u_j)}{\partial x_j} = 0 \tag{15}$$

$$\frac{\partial (\rho_m u_j)}{\partial t} + \frac{\partial (\rho_m u_i u_j)}{\partial x_j} = -\frac{\partial p}{\partial x_i} + \frac{\partial}{\partial x_i}(\mu_m \frac{\partial u_i}{\partial x_j}) \tag{16}$$

LES separates flow variables into two components: grid-scale (GS) parts that it resolves directly, and subgrid-scale (SGS) parts that it models [75]. To obtain a filtered variable $\phi$, LES convolves the unfiltered quantity with a filter function $G = G(X, \Delta)$, which acts over a characteristic length scale $\Delta$ and time scale $\tau$ [76]. It removes scales minor than these $\Delta$ and $\tau$ using equation (17), while equation (18) captures the SGS effects through a residual term.

$$\overline{\phi(x,t)} = \int_{-\infty}^{\infty} \int_{-\infty}^{\infty} \phi(r,t') . G(x-r, t-t') dt' dt \tag{17}$$

$$\phi = \bar{\phi} - \phi' \tag{18}$$

Applying the filtering operation yields the modified governing equations (19) and (20), where variables with an overbar indicate filtered quantities. The resulting SGS stress tensor $\tau_{ij}$ appears in equation (21) [77] and requires closure via a subgrid-scale model.

$$\frac{\partial \rho_m}{\partial t} + \frac{\partial (\rho_m \bar{u}_j)}{\partial x_j} = 0 \tag{19}$$

$$\frac{\partial (\rho_m \bar{u}_j)}{\partial t} + \frac{\partial (\rho_m \bar{u}_i \bar{u}_j)}{\partial x_j} = -\frac{\partial \bar{p}}{\partial x_i} + \frac{\partial}{\partial x_i}(\mu_m \frac{\partial \bar{u}_i}{\partial x_j}) - \frac{\partial \tau_{ij}}{\partial x_j} \tag{20}$$

$$\tau_{ij} \approx \rho(\overline{u_i u_j} - \bar{u}_i \bar{u}_j) \tag{21}$$

In the equations outlined previously, properties denoted by the overline symbol ( ¯ ) represent filtered variables. To model $\tau_{ij}$, a commonly adopted approach is the eddy viscosity, which relates the deviatoric part of the SGS stress to the filtered strain rate through an effective turbulent viscosity, as described in equation (22):

$$\tau_{ij} = \frac{1}{3}\tau_{kk}\delta_{ij} - \nu_t(\frac{\partial \bar{u}_i}{\partial x_j} + \frac{\partial \bar{u}_j}{\partial x_i}) \tag{22}$$

In this equation, $\nu_t$ represents the SGS eddy viscosity. The term $\tau_{kk}$ accounts for the isotropic component of the SGS stress and is absorbed into the modified pressure term. Among the various SGS models, the Smagorinsky model [78] is one of the most widely used. It estimates the SGS eddy viscosity $\nu_t$ based on the local strain rate magnitude and filter size. Equations (23) and (24) express $\nu_t$ in terms of a model constant $C_s$, the filter length $\Delta$, and the magnitude of the strain rate $|\bar{S}| = (2\bar{S}_{ij}\bar{S}_{ij})$ [79].

$$\nu_t = C_s \bar{\Delta}^2 |\bar{S}| \tag{23}$$

$$\bar{S}_{ij} = \frac{\partial \bar{u}_{ij}}{\partial x_j} + \frac{\partial \bar{u}_j}{\partial x_j} \tag{24}$$

The Smagorinsky model is well-regarded for its simplicity, reliability, and computational efficiency. It has demonstrated effectiveness in hydrofoil simulations and various engineering flows that require a focus on large-scale turbulent structures. Although more sophisticated methods like spectral LES and Direct Numerical Simulation (DNS) provide greater accuracy [80, 81], their high computational requirements can render them impractical for intricate, large-scale simulations. In comparison, using LES with the Smagorinsky model offers a sensible compromise between accuracy and efficiency, effectively capturing phenomena such as cavitation onset, bubble collapse, and vortex shedding [82].

However, this model's accuracy depends heavily on the calibration of the model constant $C_s$, which can vary based on flow configuration. Ferziger and McMillan [83] emphasized the need to tune this constant for optimal accuracy in internal flows as $C_s = 0.1$. Comprehensive descriptions can be found in Fig. 26 (Refer to Appendix A). This study implements LES using Ansys Fluent to simulate the laminar-to-turbulent transition under cavitating conditions, a method well-validated in the literature [66, 84]. Its ability to resolve unsteady turbulent flow features makes LES especially suitable for capturing transitional phenomena in cavitating environments.

**2.4. k-ω SST Turbulence Model**

This study utilizes the k-ω shear stress transport (SST) model for accurate turbulence representation in cavitating flows. Created by Menter [85], the k-ω SST formulation combines the benefits of two popular turbulence models: the k-ω model, known for its precision in near-wall areas, and the k-ε model, praised for its reliability in distant fields. This hybrid method improves the simulation of adverse pressure gradients and flow separation in intricate geometries.

The equations for turbulence kinetic energy (k) and specific dissipation rate (ω) are formulated in equations (25) and (26). They encompass production, dissipation, and diffusion terms, coupled with a blending function that varies between the k-ω and k-ε models based on the flow region.

$$\frac{\partial}{\partial t}(\rho k) + \frac{\partial}{\partial x_j}(\rho k u_j) = \frac{\partial}{\partial x_j}\left(\left(\mu + \frac{\mu_t}{\sigma_{k3}}\right)\frac{\partial k}{\partial x_j}\right) + \tau_{ij}\frac{\partial u_i}{\partial x_j} - \beta^* \rho k \omega \quad (25)$$

$$\frac{\partial(\rho\omega)}{\partial t} + \frac{\partial(\rho u_j \omega)}{\partial x_j} = \frac{\partial}{\partial x_j}\left(\left(\mu + \frac{\mu_t}{\sigma_{\omega 3}}\right)\frac{\partial \omega}{\partial x_j}\right) + \frac{\omega}{k}\left(\alpha_3 \tau_{ij}\frac{\partial u_i}{\partial x_j}\right) - \beta_3 \rho \omega^2$$
$$+ (1-F_1)2\rho \frac{1}{\omega \sigma_{\omega 2}}\frac{\partial k}{\partial x_j}\frac{\partial \omega}{\partial x_j} \quad (26)$$

The model coefficients $\alpha_3$, $\beta_3$, $\sigma_{k3}$, and $\sigma_{\omega 3}$ are blended between the k-ω and modified k-ε values using a function $F_1$, as shown in equation (27). This blending enables a smooth transition from one model to another based on the distance from the wall and local flow conditions [86-88].

$$\psi = F_1 \psi_{k\omega} + (1-F_1)\psi_{k\varepsilon,} \quad \alpha_3 = F_1 \alpha_1 + (1-F_1)\alpha_2$$
$$k-\omega: \alpha_1 = 5/9, \quad \beta_1 = 3/40, \quad \sigma_{k1} = 2, \quad \sigma_{\omega 1} = 2, \quad \beta^* = 9/100 \quad (27)$$

The eddy viscosity is calculated based on the strain rate tensor $s_{ij}$, defined in equations (28) and (29). This formulation allows the turbulence model to respond to the flow field's local shear and rotation rates.

$$\tau_{ij} = \mu_t \left(2s_{ij} - \frac{2}{3}\frac{\partial u_k}{\partial x_k}\delta_{ij}\right) - \frac{2}{3}\rho k \delta_{ij} \quad (28)$$

$$S = \sqrt{2s_{ij}s_{ij}}, \quad s_{ij} = \frac{1}{2}\left(\frac{\partial u_i}{\partial x_j} + \frac{\partial u_j}{\partial x_i}\right) \tag{29}$$

To control excessive turbulent diffusion in regions of smooth surfaces, the SST model introduces a blending function $F_1$ defined in the equation (30), where the intermediate variable $\Gamma$ is given in the equation (31). The cross-diffusion term $CD_{k\omega}$, used in this blending, is defined in the equation (32) [86, 87].

$$F_1 = \tanh(\Gamma^4) \tag{30}$$

$$\Gamma = \min\left(\max\left(\frac{\sqrt{k}}{\beta^*\omega y}; \frac{500\nu}{\omega y^2}\right); \frac{4\rho\sigma_{\omega 2}k}{CD_{k\omega}y^2}\right) \tag{31}$$

$$CD_{k\omega} = \max\left(2\rho\sigma_{\omega 2}\frac{1}{\omega}\frac{\partial k}{\partial x_j}\frac{\partial \omega}{\partial x_j}, 10^{-20}\right) \tag{32}$$

To further improve near-wall modeling, a limiter is applied to the eddy viscosity using the equation (33), where $S$ is an invariant measure of the strain rate and $F_2$ is a secondary blending function, defined in equations (34) and (35) [86-88].

$$\mu_t = \rho\frac{k}{\max(\omega, SF_2)} \tag{33}$$

$$F_2 = \tanh(\Gamma_2^2) \tag{34}$$

$$\Gamma_2 = \max\left(\frac{2\sqrt{k}}{\beta^*\omega y}, \frac{500\nu}{\omega y^2}\right) \tag{35}$$

Accurate boundary layer integration using low-Reynolds-number models is preferred but requires fine grids, which are impractical for complex geometries. Coarse grids diminish accuracy, and wall functions become unreliable with insufficient resolution. Grotjans and Menter [89] developed an automatic wall treatment that smoothly blends low-Re and wall function formulations using a y+ dependent approach to address this issue. This method, implemented with the SST model, optimally adapts to grid quality and avoids poor results on under-resolved meshes. This treatment ensures smooth transitions between the linear sublayer and logarithmic layer, as seen in equations (36) to (39), based on Menter et al. [88].

$$\omega_{\log} = \frac{1}{0.3\kappa}\frac{u_\tau}{y}; \quad \omega_{Vis} = \frac{6\upsilon}{0.075 y^2} \tag{36}$$

$$\omega_1(y^+) = \sqrt{\omega_{Vis}^2(y^+) + \omega_{\log}^2(y^+)} \tag{37}$$

$$u_\tau^{Vis} = \frac{U_1}{y^+}; \quad u_\tau^{\log} = \frac{U_1}{\frac{1}{\kappa}\ln(y^+)+c} \tag{38}$$

$$u_\tau = \sqrt[4]{(u_\tau^{Vis})^4 + (u_\tau^{\log})^4} \tag{39}$$

This formulation enables the model to automatically transition between low-Reynolds-number modeling near the wall and a wall-function approach based on local grid resolution and y+. As shown in the research by Menter et al. [88], this ensures precise turbulence representation across various flow regimes, especially those involving separation, adverse pressure gradients, and cavitation instabilities.

**2.5. Standard k-ε Turbulence Model**

The standard k-ε model ranks among the most popular turbulence models in engineering, largely because of its computational efficiency and adequate accuracy across diverse flows. It is categorized as a two-equation model and relies on the Boussinesq eddy viscosity hypothesis, which connects Reynolds stresses to the mean strain rate via an isotropic turbulent viscosity [90]. The transport equations governing k and ε are presented in equations (40) and (41), respectively.

$$\frac{\partial}{\partial t}(\rho k) + \frac{\partial}{\partial x_j}(\rho k u_j) = \frac{\partial}{\partial x_j}\left[\left(\mu + \frac{\mu_t}{\sigma_k}\right)\frac{\partial k}{\partial x_j}\right] + P_k - \rho\varepsilon \qquad (40)$$

$$\frac{\partial(\rho\varepsilon)}{\partial t} + \frac{\partial(\rho u_j \varepsilon)}{\partial x_j} = \frac{\partial}{\partial x_j}\left[\left(\mu + \frac{\mu_t}{\sigma_\varepsilon}\right)\frac{\partial \varepsilon}{\partial x_j}\right] + C_{1\varepsilon}\frac{\varepsilon}{k}P_k - C_{2\varepsilon}\rho\frac{\varepsilon^2}{k} \qquad (41)$$

Here, $\rho$ is the fluid density, $u_j$ is the velocity component, and $\mu$ is the dynamic viscosity. The production term $P_k$, representing the generation of turbulent kinetic energy, is computed using the equation (42):

$$P_k = \mu_t\left(\frac{\partial u_i}{\partial x_j} + \frac{\partial u_j}{\partial x_i}\right)\frac{\partial u_i}{\partial x_j} \qquad (42)$$

The turbulent or eddy viscosity $\mu_t$, essential for modeling momentum diffusion in turbulent flows, is determined from the equation (43):

$$\mu_t = \rho C_\mu \frac{k^2}{\varepsilon} \qquad (43)$$

The model relies on empirically calibrated constants, which are well-established for high-Reynolds-number wall-bounded flows [90, 91]:

$$C_\mu = 0.09, \quad C_{1\varepsilon} = 1.44, \quad C_{2\varepsilon} = 1.92, \quad \sigma_k = 1.0, \quad \sigma_\varepsilon = 1.3 \qquad (44)$$

While the model operates under the assumption of isotropic turbulence, it performs effectively in fully developed turbulent flows, particularly away from boundaries. In cases of cavitating flows, the standard k-ε model produces dependable predictions when integrated with volume fraction transport equations and a homogeneous mixture formulation. Park and Rhee [92] illustrated that the model accurately represents pressure variations caused by turbulence and improves cavity closure by addressing turbulent diffusion at the cavity interface.

While the standard k-ε model has its strengths, it shows limitations in situations characterized by strong streamline curvature, flow separation, or swirl, particularly where anisotropic turbulence prevails. For enhanced accuracy, advanced models like the realizable k-ε model or the k-ω SST model are suggested.

### 2.5.1 Realizable k-ε Turbulence Model

The realizable k-ε model, developed by Shih et al. [93], enhances the traditional k-ε formulation by tackling important physical and mathematical shortcomings. The term "realizable" indicates that this model guarantees the computed Reynolds stress tensor is physically consistent, meaning it adheres to constraints like the positivity of normal stresses and the Schwarz inequality for shear stresses.

This enhanced model introduces two fundamental modifications:

1. A new transport equation for the dissipation rate $\varepsilon$, replacing the standard form.

2. A redefined formulation for the eddy viscosity $\mu_t$, in which the coefficient $C_\mu$ is no longer constant but dynamically determined based on the local flow field.

In the realizable model, the turbulent (eddy) viscosity is given by:

$$\mu_t = \rho C_\mu \frac{k^2}{\varepsilon} \tag{45}$$

Unlike the standard k-ε model, the coefficient $C_\mu$ in the realizable version varies with the flow and is computed as [93-95]:

$$C_\mu = \cfrac{1}{A_0 + A_s \cfrac{U^* \varepsilon}{k}} \tag{46}$$

Here, $A_0 = 4.04$, while $A_s$ is a function of the local strain rate, and $U^*$ is a velocity scale derived from the mean strain rate tensor. This flow-dependent formulation enables the model to adapt to zones of high deformation, strong curvature, or rotational motion, improving prediction accuracy in complex turbulent regimes.

The transport equation for $\varepsilon$ is also redefined to capture turbulence dissipation under varying flow conditions better:

$$\frac{\partial(\rho\varepsilon)}{\partial t} + \frac{\partial(\rho u_j \varepsilon)}{\partial x_j} = \frac{\partial}{\partial x_j}\left(\left(\mu + \frac{\mu_t}{\sigma_\varepsilon}\right)\frac{\partial \varepsilon}{\partial x_j}\right) + \rho C_1 S \varepsilon - C_2 \rho \frac{\varepsilon^2}{k} + \upsilon_\varepsilon + C_1 \frac{\varepsilon}{k} C_k \tag{47}$$

In this equation, $S$ represents the modulus of the mean strain rate tensor, and the coefficients $C_1$, $C_2$, and $C_3$ are dynamically updated based on local flow characteristics, enhancing the model's responsiveness.

The realizable k-ε model has demonstrated superior performance in various challenging flow scenarios, including boundary layer separation, swirling and rotating flows, recirculating regions, and free shear layers.

## 3. Numerical strategy

### 3.1. Porous hydrofoil case: Discretization and geometry

This numerical study examines the behavior of porous media in relation to cavitating flow over a NACA 66 (MOD) hydrofoil. The simulation utilizes the Pressure–Implicit with Splitting of Operators (PISO) algorithm to ensure an accurate coupling of pressure and velocity and employs the Pressure Staggering Option (PRESTO) scheme for pressure discretization. A second-order upwind differencing method is consistently applied across the domain to enhance the spatial accuracy of convective fluxes.

This study conducted a time-step sensitivity analysis using values of 0.05 s, 0.001 s, and 0.0001 s to evaluate simulation accuracy and efficiency. As illustrated in Fig. 1, a time step of 0.001 s provided the optimal balance. A combined assessment of numerical stability guided this choice, along with the ability to resolve unsteady cavitation features and maintain computational efficiency. Since the SST k-ω and the realizable k-ε turbulence models exhibit very similar near-wall and transient flow behaviors, we utilized a time step of 0.001 s for the k-ε simulations, as this duration has already been validated for the LES and SST k-ω cases. This consistency ensures the same step across all turbulence models, facilitating valid comparisons. This analysis assesses solution stability and convergence across multiple turbulence models. The optimal time step is chosen based on minimizing variations in key flow quantities, such as cavity length and lift coefficient, which indicate temporal resolution independence.

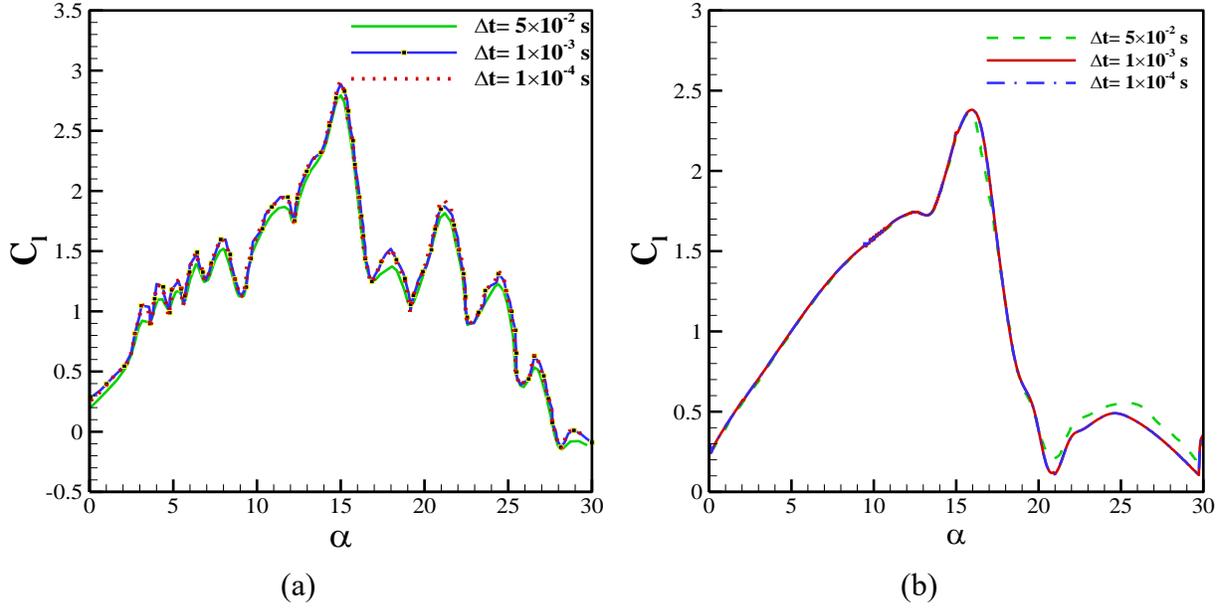

Fig. 1. Time step independence analysis for the lift coefficient at a cavitation number of $\sigma = 8$, evaluated using two different turbulence models: (a) LES and (b) k-ω SST.

The VOF method addresses the interface between vapor and liquid phases. It operates under the assumption that these two phases are immiscible and possess identical velocity and temperature fields throughout a clearly defined interface. Additionally, it achieves second-order accuracy in both spatial and temporal discretization. Table 1 presents a detailed summary of all pertinent simulation parameters.

Table 1. Summary of simulation configuration and model parameters

| Algorithm | Time Step (s) | Solution Accuracy | Turbulence Model | Two-Phase Model | Cavitation Model | Angle of Attack (AOA) |
|---|---|---|---|---|---|---|
| PISO | 0.001 for 2D case 0.0001 for 3D case | second-order | LES, k-ω SST, k-ε | VOF | Schnerr-Sauer | 9° for all cases |

The simulation is conducted at a Reynolds number of $Re = 6.58 \times 10^5$, using the hydrofoil chord length as the characteristic length. Fig. 2(a) shows the computational domain that spans 5.5c upstream and 10.5c downstream from the chord midpoint, forming a two-dimensional mesh of 236,787 cells. A combination of structured and unstructured meshing is employed, as depicted in Fig. 2(b). The hydrofoil surface is defined as a no-slip boundary condition. Fig. 3(a) presents the computational domain for the three-dimensional case, which extends 5c upstream and 11c downstream from the leading edge of the hydrofoil, resulting in a mesh consisting of 2,325,220 cells. Like the two-dimensional simulation, structured and unstructured meshing enhances accuracy and computational efficiency. Additionally, a spanwise length of 0.3c is defined in the z-direction, as shown in Fig. 3(b). This setup enables the precise resolution of three-dimensional cavitation structures, including tip effects, critical for capturing the full complexity of the unsteady cavitating flow.

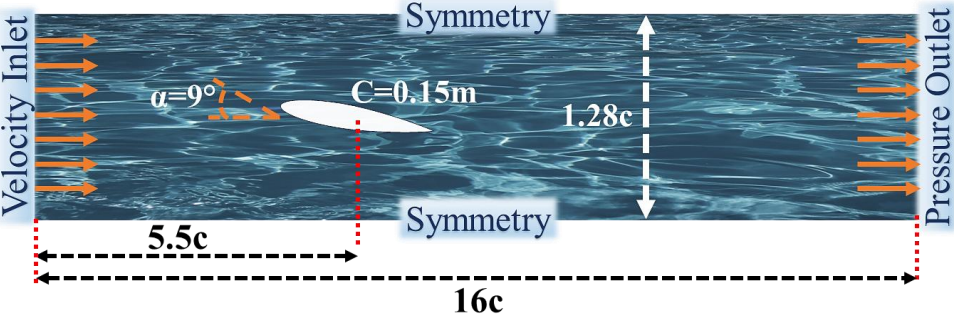

(a)

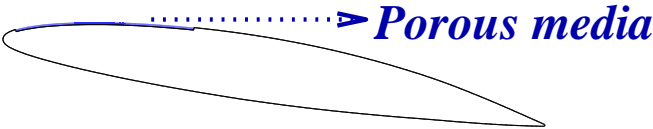

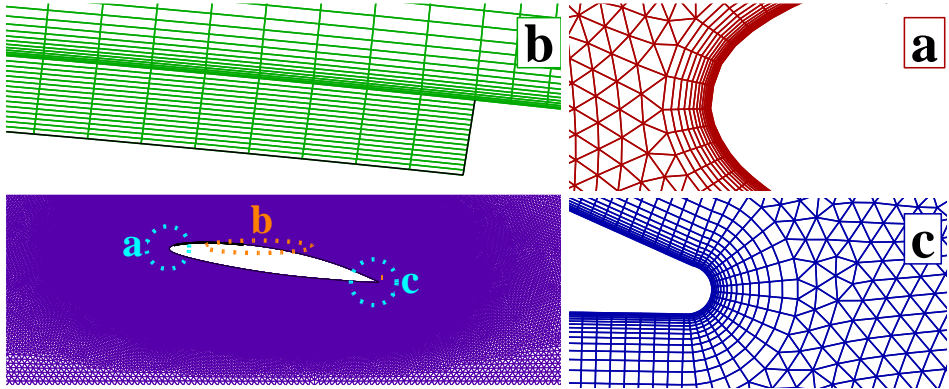

(b)

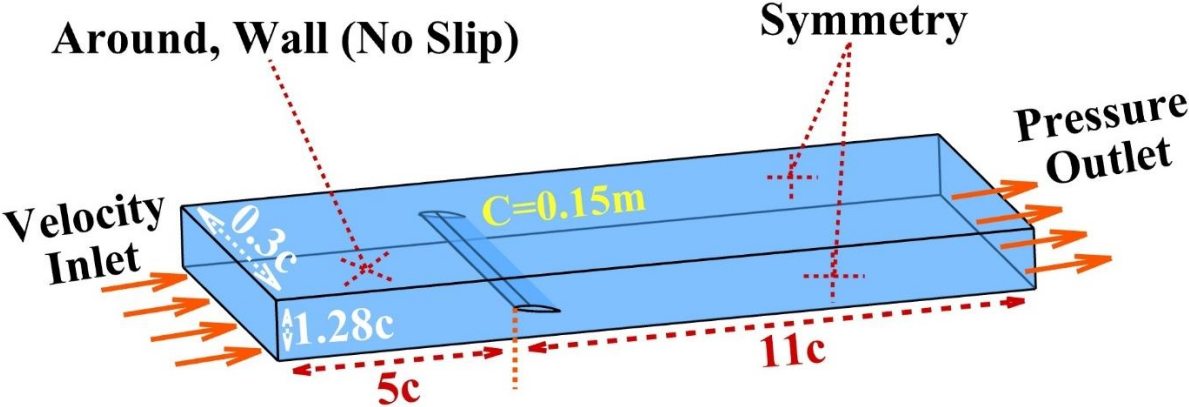

Fig. 2. NACA 66 (MOD) hydrofoil setup: (a) Two-dimensional computational domain; (b) Grid topology

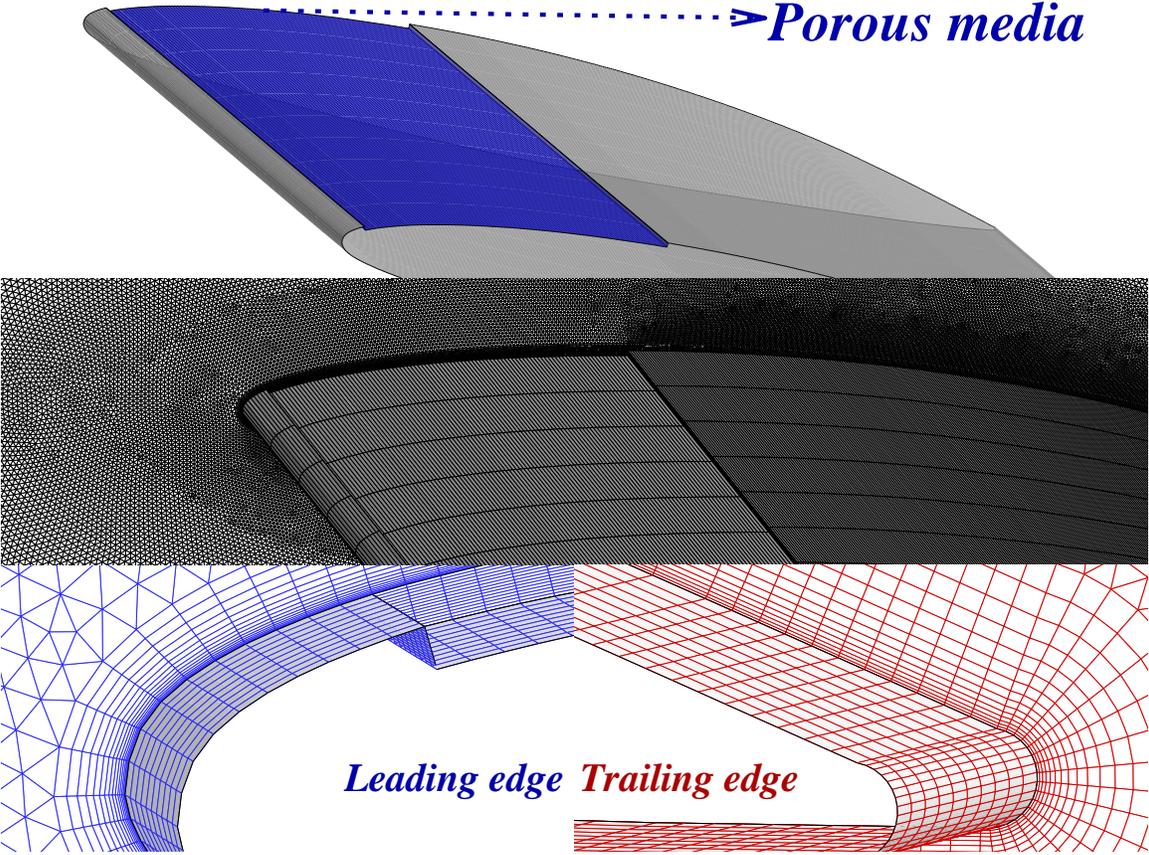

Fig. 3. NACA 66 (MOD) hydrofoil setup: (a) Three-dimensional computational domain; (b) Grid topology

Boundary conditions are derived and validated against experimental data to ensure physical reliability. The cavitation number σ is defined by the equation $\sigma = P_\infty - P_g / 0.5\rho V_\infty^2$, where $P_\infty = 103088 \text{Pa}$ is the reference static pressure, $P_v$ is the vapor pressure, $\rho$ is the liquid density, and $V_\infty = 5 m/s$ is the inlet velocity. The fluid is assumed to be pure water at 25 °C, with physical properties assigned as follows: $\rho_l = 999.19 kg/m^3$, $\mu_l = \rho_l v_l = 1.139 \times 10^{-3} Pa.s$, $\rho_v = 0.02308 kg/m^3$, and $\mu_v = 9.8626 \times 10^{-3} Pa.s$. Simulations are performed for two cavitation numbers, $\sigma = 0.7$ and $\sigma = 1$, corresponding to 94345.08Pa and 90598.13Pa vapor pressures, respectively.

The porous area is represented as a momentum sink through Ergun's equation, which is utilized in ANSYS Fluent to depict flow resistance in porous media semantically. The governing relationship is given as:

$$\frac{\Delta p}{L} = \frac{150\mu_m}{D_p^2}\frac{(1-\phi)^2}{\phi^3}V_\infty + \frac{1.75\rho_m}{D_p}\frac{(1-\phi)}{\phi^3}V_\infty^2 \tag{48}$$

Where $D_p$ is the mean particle diameter, set to 0.1 mm, and $\phi$ is the porosity, defined by equation $\phi = 1 - V_{solid}/V_{shape}$. The porosity is fixed at 0.95, matching the value used in the corresponding experimental measurements [41, 96]. The porous zone is defined along the suction side of the hydrofoil, extending over a length of 0.35m (1/3c) and a depth of 0.005m, consistent across both 2D and 3D cases.

### 3.2. Grid independence analysis

Table 2 compiles comprehensive mesh values, and Fig. 4 presents the grid independence analysis for the LES turbulence model. Our results show excellent consistency, highlighting the accuracy and effectiveness of the chosen meshing technique despite having fewer total mesh cells than the reference study [97]. This finding is crucial as our simulation significantly reduces computational costs while achieving similar predictive performance. Although we use second-order accuracy for enhanced solution authenticity, the referenced work [97] utilized a finer grid with first-order discretization techniques. This study benefits from faster convergence and shorter solution times by employing the PISO algorithm. Observations of improved stability arise from combining a lower mesh count, the PISO method, and a high-order numerical scheme. Collectively, these approaches boost the efficiency and robustness of the CFD simulations.

Regarding the lift coefficient, Grid 3 aligns with Grid 4, as shown in Fig. 4(a), affirming their suitability for the main simulation. In addition to demonstrating grid convergence, near-wall resolution is vital, particularly for modeling wall-bounded turbulent structures and vortex dynamics. This study emphasizes the dimensionless wall distance, $y^+$, critical for accurately capturing wall-bounded turbulence as it integrates the dynamic Smagorinsky subgrid-scale model within an LES framework. A standard criterion in turbulence modeling, the $y^+$ value indicates the non-dimensional distance from the wall with respect to the viscous sublayer thickness. Precise resolution of the near-wall region is essential, especially in LES, where excessive $y^+$ values due to inadequate grid refinement can impair the Smagorinsky model's performance and lead to inaccuracies in predicting wall shear and vortex shedding occurrences [98-100]. The grid was meticulously constructed to maintain $y^+ < 1$ throughout all simulation stages, allowing for the direct resolution of the viscous sublayer and enhancing the accuracy of boundary layer predictions, thus ensuring high-fidelity results. Additionally, fifteen prism layers were implemented next to the hydrofoil surface to provide sufficient refinement to meet the near-wall resolution requirements. The congruence between the numerical results and experimental data [101] further reinforces the appropriateness of the selected grid.

The same mesh was utilized for the k-ω SST and k-ε turbulence models, although mesh independence was specifically verified for the LES model. The LES method's ability to capture a wide range of turbulent scales, particularly near walls and in shear layers, typically requires finer grids. As a result, the mesh used for LES adheres to stricter spatial resolution standards than those generally necessary for RANS-based models like k-ω SST and k-ε, which rely on turbulence closure without directly resolving the small-scale eddies. Therefore, the LES mesh provides adequate accuracy for the RANS models, ensuring consistent predictions without additional mesh independence analyses for each method.

Table 2. Multiple mesh settings for two two-dimensional models. The list indicates the number of grid cells employed in the vertical and streamwise orientations.

| $\sigma = 8$ | $n_\eta$ nodes | $n_\xi$ nodes | Cells |
|---|---|---|---|
| Grid 1 (Coarse) | 30 | 300 | 63,096 |
| Grid 2 (Medium) | 43 | 425 | 121,874 |
| Grid 3 (Fine) | 61 | 600 | 236,787 |
| Grid 4 (Finer) | 86 | 850 | 461735 |

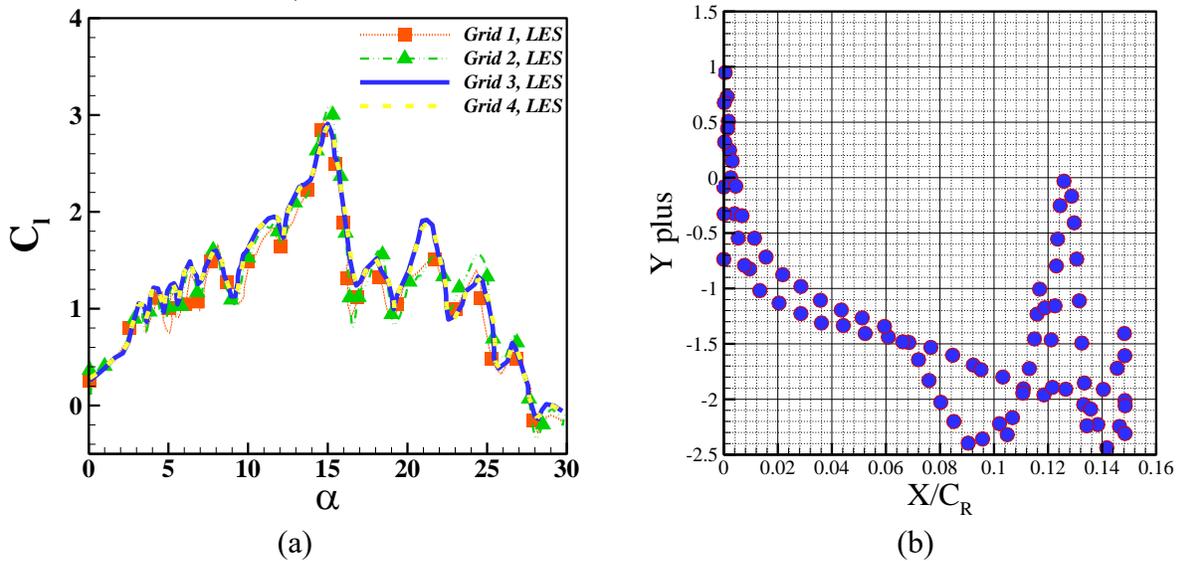

(a)   (b)

Fig. 4. Examination of several grid layouts on the NACA 66 (MOD) hydrofoil: (a) Grid independence, (b) Average cell distance from the wall ($y^+$) for the ideal mesh

We did not repeat mesh and time-step independent studies in the current three-dimensional simulations. This decision was based on the findings and conclusions from the previous two- and three-dimensional studies [102], which identified the ideal hydrofoil geometry and demonstrated remarkable agreement with experimental data. That validated study guided the development of the three-dimensional configuration used in this work. We followed the same meshing strategy, maintaining consistent near-wall resolution (e.g., $y^+ < 1$), applying equivalent refinement in critical flow regions such as the suction side, and ensuring similar cell distribution near the hydrofoil surface.

Since the LES turbulence model requires the most stringent grid resolution, and these needs were met and confirmed in the prior study, the chosen mesh is deemed adequate for accurate three-dimensional LES results. Furthermore, there is no notable difference in the flow physics or numerical methods between the existing and prior setups. The agreement between the flow behavior in two and three dimensions, as later demonstrated in Fig. 8 and Fig. 9, reinforces the reliability and precision of the current mesh. For these reasons, and to prevent unnecessary computational expenses, conducting further mesh or time step independence tests is considered unnecessary for this simulation.

### 3.3. Validation of cavitation simulation

The k-ω SST turbulence model was employed for simulation validation due to its well-documented ability to provide accurate predictions both near the wall and in the free stream, as it combines the strengths of the k-ω SST and k-ε models. Evaluated for two oscillation rates, $\alpha = 63°/s$ and $\alpha = 6°/s$, Fig. 5 illustrates the lift, drag, and momentum coefficients at a cavitation number $\sigma = 8$ for validation reasons. Experimental observations from Ducoin et al. [101], experimental results presented in Huang et al. [97], and experimental–numerical data from Ducoin et al. [103] show good agreement among the results. In contrast, the middle and right images show the anticipated vapor volume percentages from the k-ω SST and 2D LES models, while the left image in Fig. 6 illustrates the cavitation patterns observed in experiments. The numerical results align perfectly with the observed data regarding the overall shape and geographical distribution of the cavitation structures.

The left image in Fig. 7 demonstrates experimental visualizations, while the center and right images represent vapor fractions generated from the 3D LES model, visualized through iso-surfaces and slices. The simulations effectively replicate the experimental cavitation structures; the 3D LES model captures finer details than its 2D counterpart. This comparison highlights enhanced accuracy and resolution, which are achieved by using three-dimensional models to depict complex cavitating flow phenomena. Strong agreement between the numerical simulations and experimental measurements indicates that the turbulence and cavitation models are generally reliable. Emphasizing the predictive accuracy of the computational framework for analyzing cavitation dynamics, the color contours in the simulations represent the shape and location of the cavitating regions observed in the monochrome experimental images.

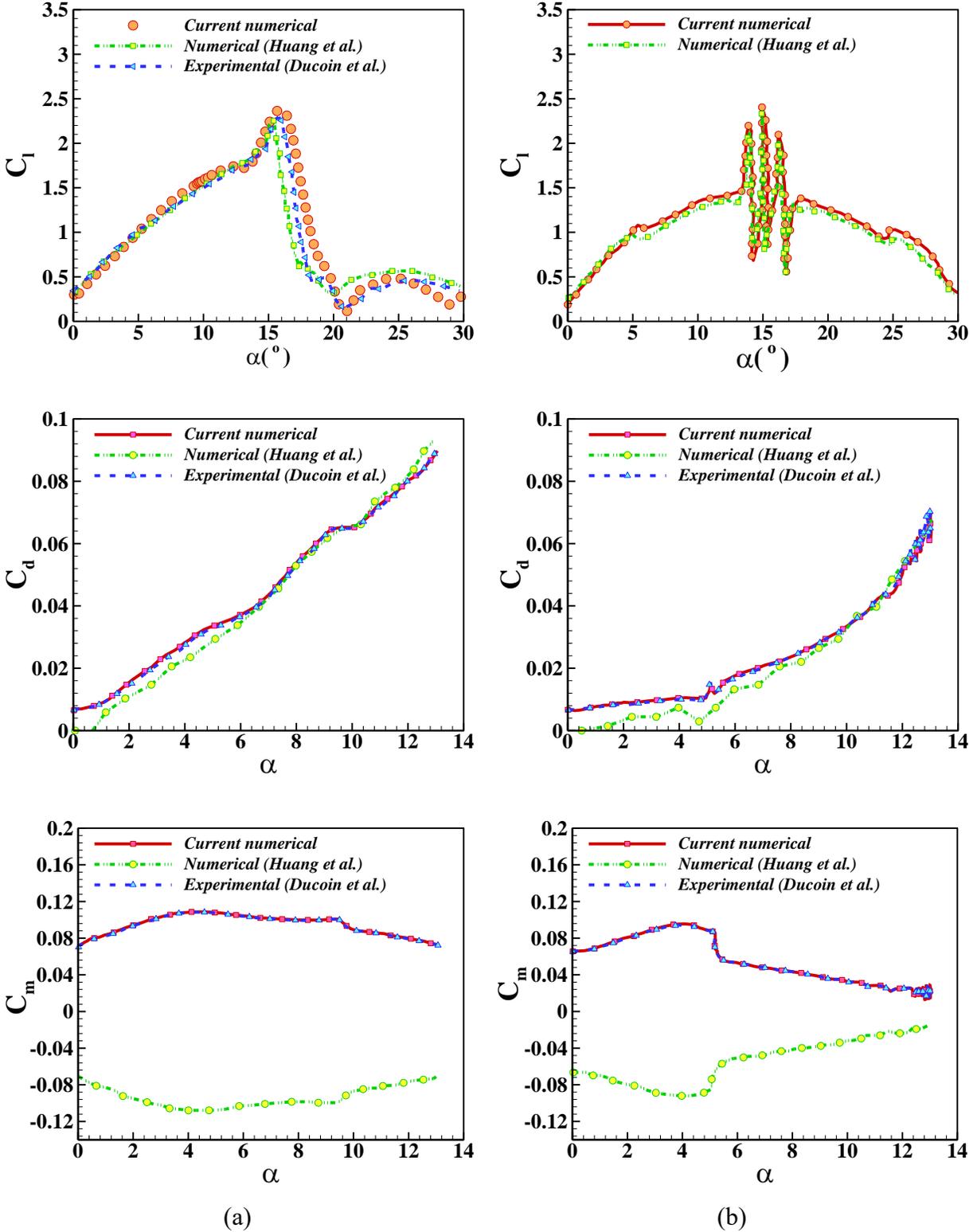

(a)           (b)

Fig. 5. Lift, drag, and momentum coefficients at cavitation number of 8 for two oscillation rates: (a) $\alpha = 63°/s$ and (b) $\alpha = 6°/s$.

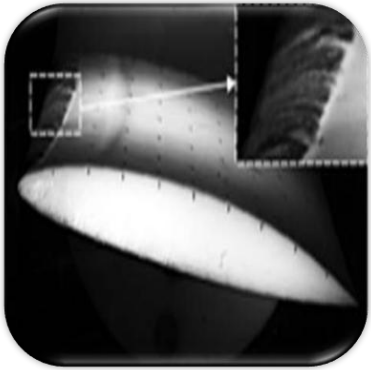 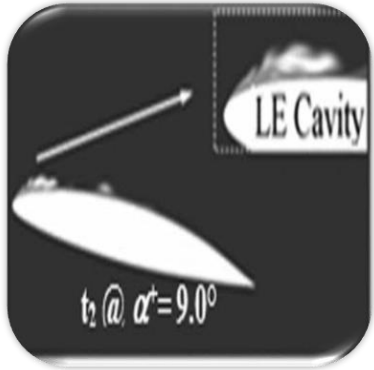 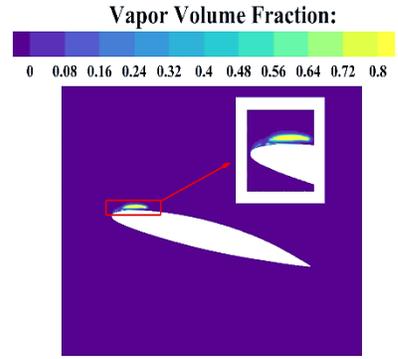

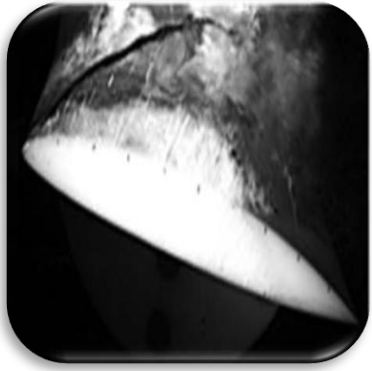 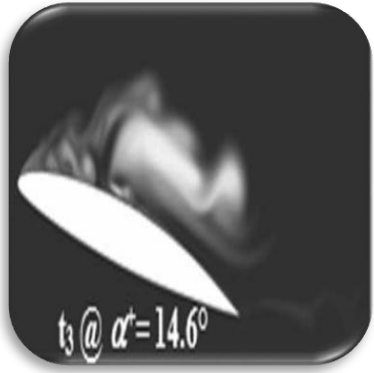 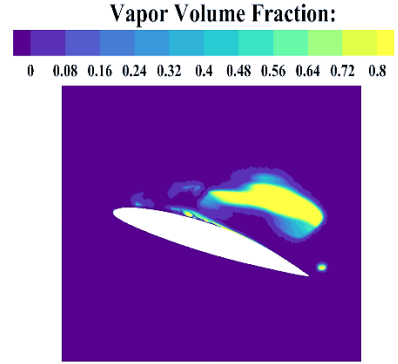

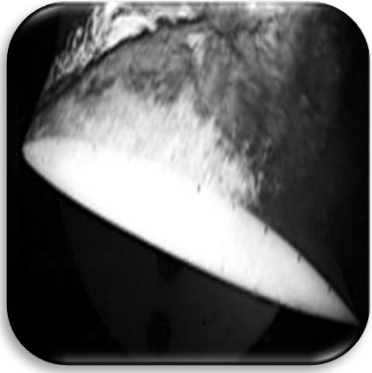 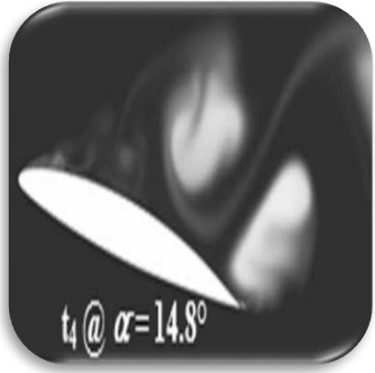 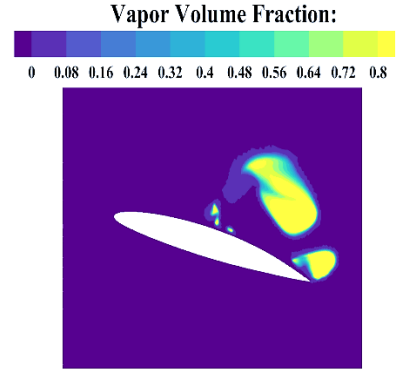

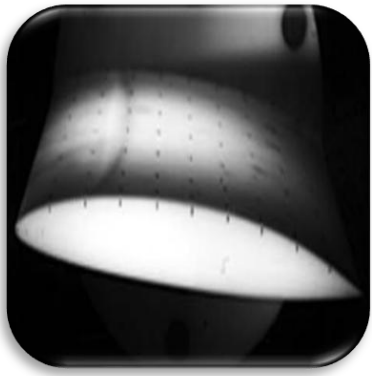 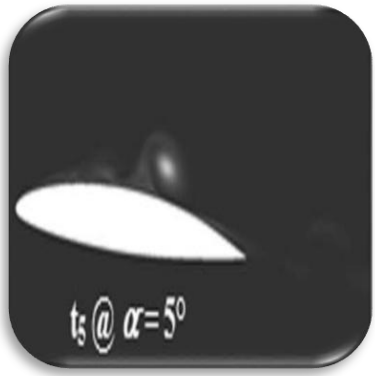 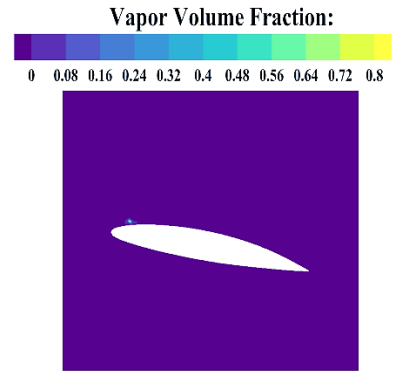

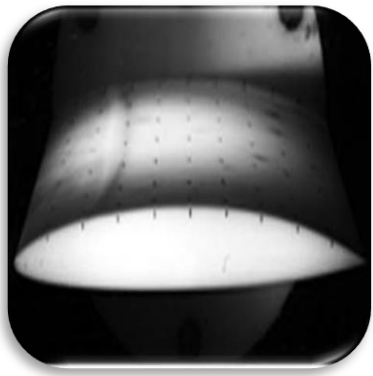 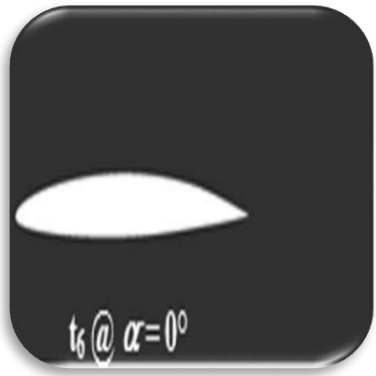 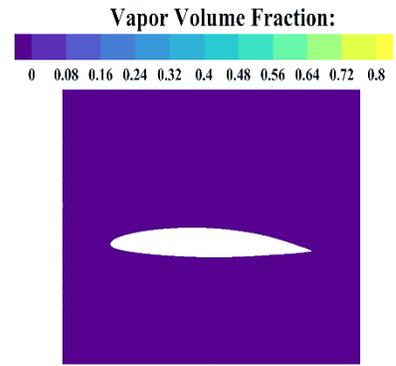

Experimental [97]　　　　　Numerical [97] (K-ω SST)　　　Current numerical (2D LES)

Fig. 6. Detailed cavitation patterns (left), expected vapor fraction utilizing the k-ω SST model (middle), and predicted vapor fraction derived from the two-dimensional LES model (right) at $\sigma = 3$ and $\alpha = 63°/s$.

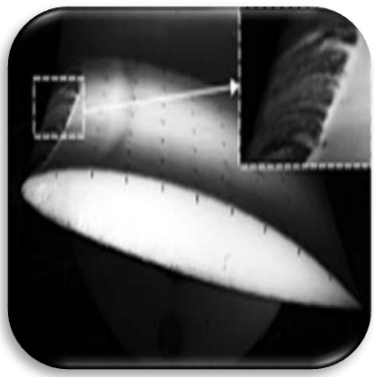 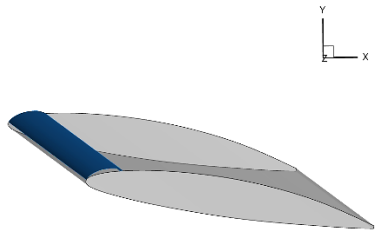 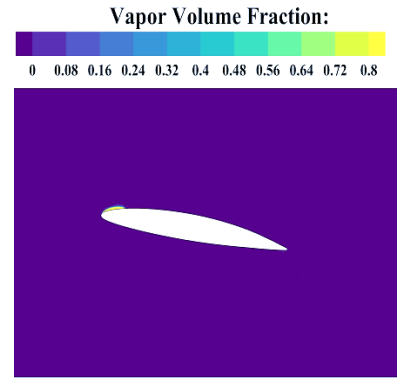

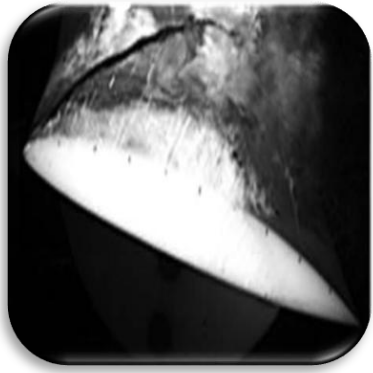 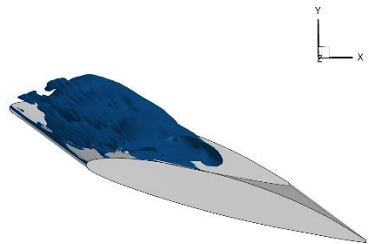 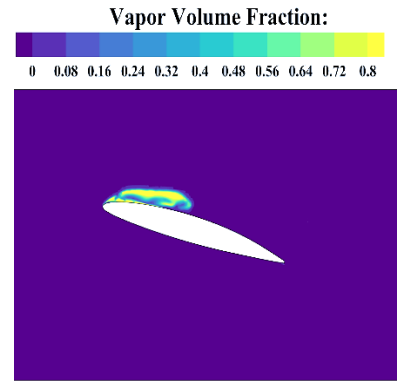

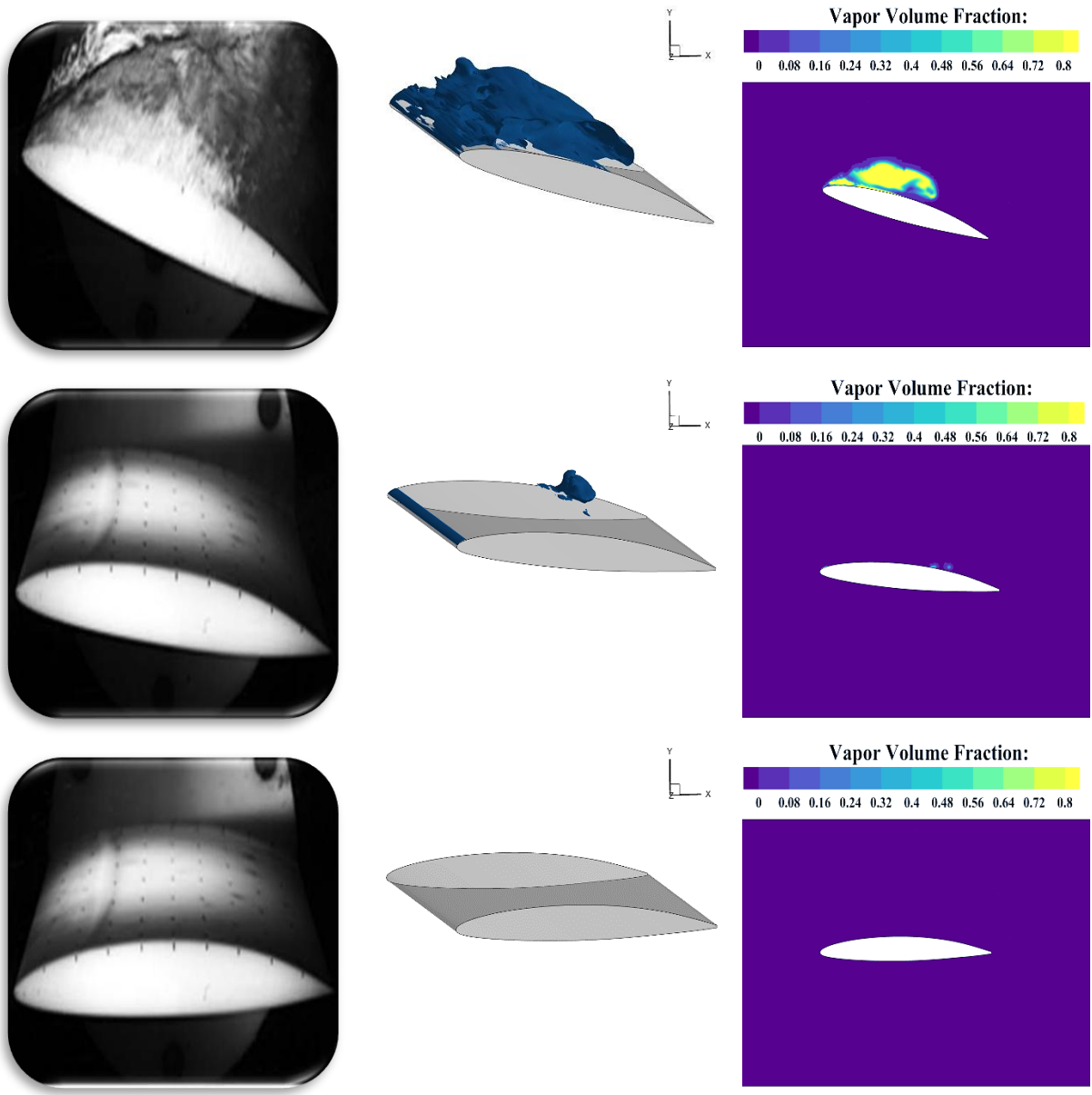

Experimental [97]          Iso Surface          Current numerical (Slice)

Fig. 7. Detailed cavitation patterns (left), iso-surface visualization (middle), and slice view (right) of vapor fraction obtained from the three-dimensional LES model at $\sigma = 3$ and $\dot{\alpha} = 63°/S$.

## 4. Results overview and analytical investigation

### 4.1. Consistency Between two- and three-dimensional simulations

Fig. 8 compares the vapor volume fraction fields for two- and three-dimensional LES simulations at a cavitation number of $\sigma = 1$. This comparison is essential in evaluating the suitability of previously validated two-dimensional results to inform the setup and expectations for the three-dimensional simulation. In the 2D LES scenario, we observe a clear instance of cavity formation and shedding. This process begins with the cavity forming at the hydrofoil's leading edge and subsequently progressing along the suction side. Behind this cavity, a re-entrant jet develops, resulting in flow separation. After this separation, vapor formation briefly resumes until a second interaction with a re-entrant jet causes the shedding of a cavity cloud into the wake. This sequence of growth, detachment, and collapse illustrates the dynamic behavior of unsteady cavitation evident in the simulation.

The 3D LES result exhibits the same fundamental flow behavior. At the beginning of the hydrofoil, vapor forms along the suction side, followed by cavity separation due to a re-entrant jet. A new cavity begins to develop again, and toward the trailing edge, a second re-entrant jet causes additional separation and generates a cavity cloud. While the 3D contours reveal added complexity, such as twisted and three-dimensional vapor filaments, the cavity growth and collapse mechanism remains consistent with that observed in the 2D simulation. The spatial evolution, shedding location, and dynamics of re-entrant jets are aligned in both dimensions. Implementing 3D structures produces more geometric variability spanwise, but overall, the cavity evolution and flow separation trends are similar. The 2D simulation captures the essential vapor dynamics and structural interferences dominating the cavitation process, precisely representing the 3D case. To further confirm this consistency, Fig. 9 compares the pressure coefficient distribution across the hydrofoil surface for 2D and 3D LES at $\sigma = 1$. The middle subfigure shows a slice taken exactly at the symmetry plane (Z=0), and the right subfigure illustrates a spanwise section at approximately 40% of the hydrofoil width. From both 3D perspectives, the pressure field again tends towards the 2D, resulting in a depression on the suction side, adverse pressure recovery, and a rise in the position of the pressure relating to each cavity's collapse. The similarities shown have demonstrated general similarities to the two-dimensional and three-dimensional solutions concerning the flow structure around several planes.

Given that LES delivers the highest resolution of unsteady and three-dimensional flow structures, and aligns closely in time step sensitivity with the k-ω SST model for 2D simulations, we performed a full 3D simulation solely for the LES case. The validated correlation between 2D and 3D LES demonstrates that the essential flow physics are accurately captured, which supports using 2D domains for the k-ε and k-ω SST models, as they are less sensitive to fine-scale structures and have lower resolution demands. Therefore, we will conduct all further simulations in two dimensions and base our findings on the validated 2D framework. The physical consistency seen between the 2D and 3D LES results validates this approach, ensuring that the accuracy and relevance of the conclusions remain intact.

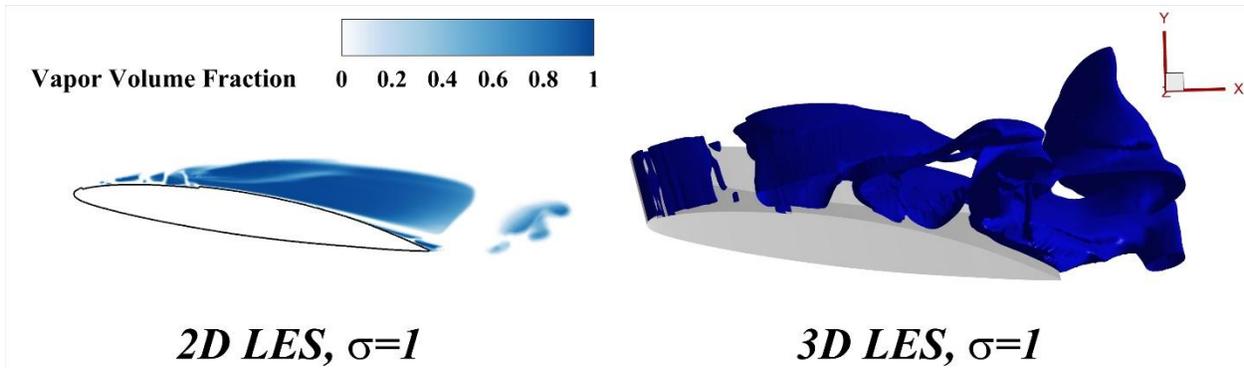

Fig. 8. Comparison of cavity formation on 2D and 3D vapor volume fraction contours on the porous hydrofoil at t=1.8s and $\sigma = 1$ with LES turbulence model.

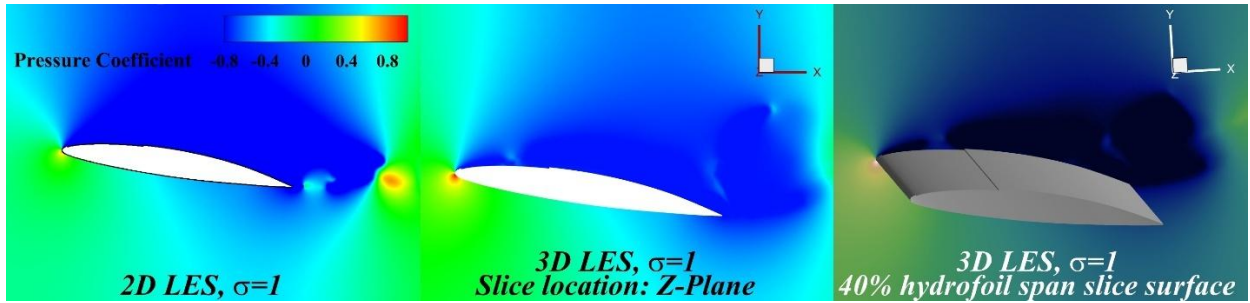

Fig. 9. Comparison of pressure coefficient contours on the porous hydrofoil using the LES turbulence model at t=1.8 s and $\sigma =1$. Left: 2D LES simulation result; middle: slice at the mid-span plane (Z=0) from the 3D LES simulation; right: surface slice at approximately 40% of the hydrofoil span from the 3D LES result.

### 4.2. Time-averaged analysis of key flow parameters

Using the LES approach, Fig. 10 clearly shows the mean pressure coefficient over one cavitation cycle, at cavitation number 1. Comparing the mean pressure coefficient between the porous and non-porous cases shows that using a porous layer on the hydrofoil surface produces a slight increase in the pressure coefficient and greater stability in terms of the flow wrapping around the hydrofoil. Fig. 10 also indicates more significant fluctuations in the non-porous case on the separated side of the hydrofoil. All of this supports the inference that persistent cavitation bubble collapses in the non-porous case produced the sudden variations in the pressure coefficient. Thus, with the use of the LES methodology, as discussed later in the thesis, it has been shown here that using porous material improves the flow stability overall.

Utilizing the LES approach, the mean pressure coefficient is clearly illustrated across one cavitation cycle at a cavitation number of 1. A comparison between the porous and non-porous cases reveals that the application of a porous layer on the hydrofoil surface leads to a slight increase in the pressure coefficient and enhances the stability of the flow as it wraps around the hydrofoil. Furthermore, the data demonstrate that the non-porous case experiences more substantial fluctuations on the separated side of the hydrofoil. This observation further supports the conclusion that the abrupt changes in the pressure coefficient resulted from the persistent collapse of cavitation bubbles in the non-porous case. Therefore, as detailed later in the thesis, the application of the LES methodology confirms that utilizing porous materials significantly enhances overall flow stability.

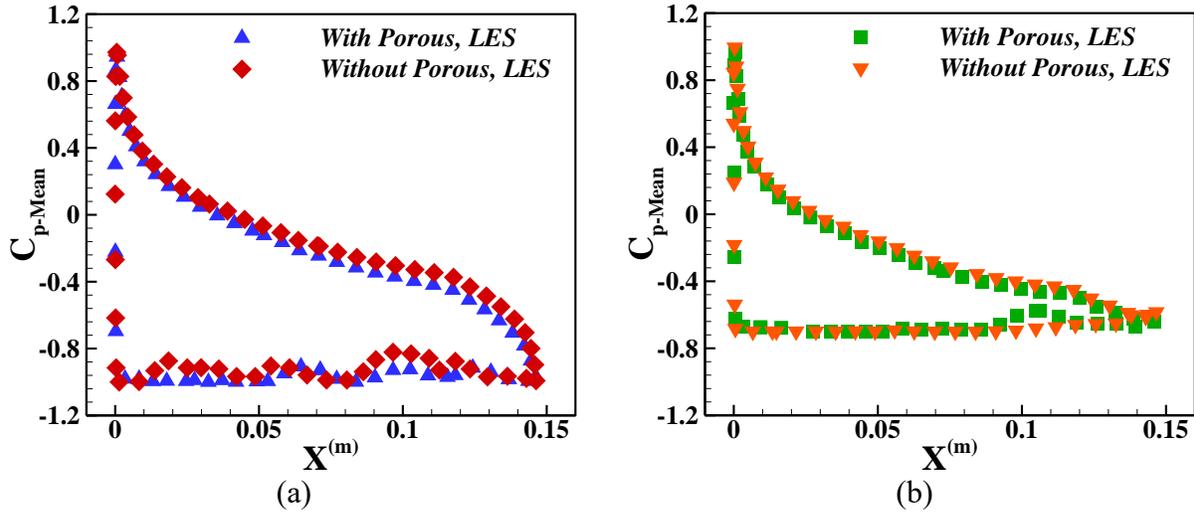

Fig. 10. Analysis of hydrofoil with and without porous media on average pressure coefficient at cavitation numbers of (a) 1, and (b) 0.7.

Comparing pressure coefficient diagrams across different turbulence models for cavitation numbers 1 and 0.7 in Fig. 11, a porous layer is shown to reduce fluctuations and improve flow stability. On the suction surface, between x = 0.07 and 0.1, the k-ε model exhibits a sharp rise in pressure, suggesting that the porous layer reduces cavitation bubble formation. This promotes flow reattachment and accelerates reconnection of the flow to the hydrofoil surface, thereby increasing the pressure coefficient. A similar trend is visible with the k-ω SST model. At x = 0.05, the k-ω SST model shows a sudden fluctuation in pressure, which is likely related to the transition of the boundary layer from laminar to turbulent. This transition typically leads to increased friction and pressure drop. Overall, in both turbulence models, the pressure coefficients are higher than those observed in the LES model, primarily due to reduced cavitation formation. This behavior is also visible in the vapor volume fraction contours (Fig. 15 and Fig. 18).

The k-ε model tends to overpredict flow reattachment and pressure recovery compared to the k-ω SST model, which provides a more gradual and accurate response. This discrepancy is well-documented and results from the over-diffusive behavior of the k-ε model in separated and adverse pressure gradient flows. Therefore, the k-ω SST model is generally more reliable in wall-bounded and transitional flow regimes. While LES can capture finer, localized flow phenomena, it has a significantly higher computational cost.

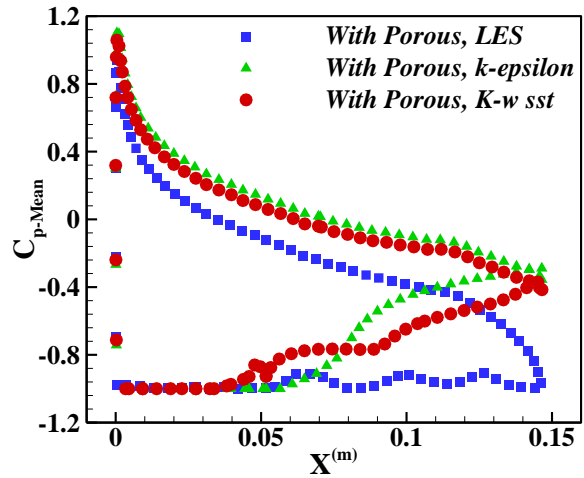

Fig. 11. Analysis of porous media on average pressure coefficient at cavitation number of 1 for various turbulence models.

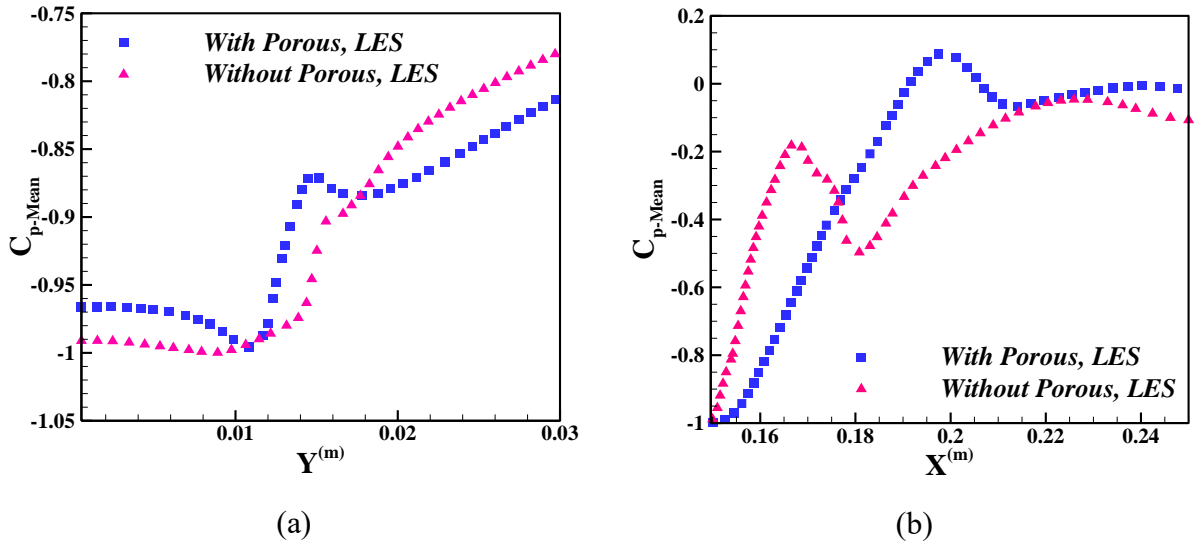

(a)            (b)

Fig. 12. Investigation of hydrofoil with and without porous media on average pressure coefficient in LES turbulence model for: (a) Along a vertical line at the mid-chord of the hydrofoil in the y-direction, and (b) Along a horizontal line near the trailing edge of the hydrofoil in the x-direction.

Fig. 12(a) illustrates the average pressure coefficient along a vertical line at the mid-point of the hydrofoil chord, and along the y-axis (y = 0.003 to 0.03m). The same average pressure coefficient is shown in Fig. 12(b) along a horizontal line that runs close to the trailing edge of the hydrofoil (x = 0.15 to 0.25m). In Fig. 6(a), (b), since the average pressure coefficient produced by a porous surface will typically exhibit less coherent cavitation bubbles produced from the porous material and the shedding behavior produced from bubble on the surface of the hydrofoil is adequately extracted from the LES turbulence model, the following plots are a valuable verification process in understanding the flow field surrounding the hydrofoil.

As presented in Fig. 12(b), the porous layer resulted in the hydrofoil delaying pressure recovery and decreasing pressure gradients to alleviate pressure spikes, which generally induce aggressive cloud cavitation, unlike the non-porous hydrofoil that developed a chaotic wake with indications of unsteady vortex shedding that would cause more noise and vibrations. Notably, substantial pressure gradients are visible immediately downstream of the trailing edge.

This behavior is summarized in Table 3, corresponding to the findings illustrated in Fig. 12.

Table 3. Comparative effects of porous media on cavitating flow characteristics using LES based on Fig. 12.

| Aspect | WITH POROUS, LES | WITHOUT POROUS, LES |
|---|---|---|
| Cavitation Shedding | Delayed and more controlled | Aggressive, unsteady cloud shedding |
| Wake Stability | Smoother wake with less turbulent dissipation | Chaotic wake and high-frequency pressure fluctuations |
| Pressure Recovery | Gradual, beneficial for lift/drag control | Sharp, potentially damaging pressure pulses |

Employing porous media is consistent with contemporary passive flow control strategies that reduce the negative impacts of cavitation, especially in LES-resolved turbulent flows. The porous layer plays a vital role in inhibiting unsteady cavity shedding and stabilizing pressure fluctuations.

Fig. 12(a) further supports the idea that porous media promotes a stable vapor layer adjacent to the hydrofoil surface. The porous material mitigates extreme negative pressures, thus lessening the severity of cavitation inception. In contrast, the non-porous model exhibits stronger suction effects and a greater local vapor density, signaling heightened cavitation activity. Furthermore, in the case of the porous medium, the wall partially absorbs the pressure wave, which helps to diminish turbulence at the vapor-liquid interface.

Table 4 summarizes these observations by concisely analyzing the data extracted from Fig. 12(a).

Table 4. Summarized comparative effects of porous media on streamwise pressure coefficient profiles along the vertical line at the mid-chord of the hydrofoil using LES.

| Region (Y) | WITH POROUS, LES | WITHOUT POROUS, LES |
|---|---|---|
| 0.00–0.01m (Near wall) | Milder suction (~ –0.96), stable film | Stronger suction (~ –1.01), aggressive inception |
| 0.01–0.018m (Transition zone) | Rapid, smooth recovery | Fluctuating, diffuse pressure increase |
| 0.018–0.03m (Outer cavity) | Linear rise, stabilized wake pressure | Steeper rise, signs of instability/collapse |

## 4.3 Analysis of the Fast Fourier Transform of hydrodynamic coefficients in an unsteady flow field

Fig. 13 presents the Fast Fourier Transform (FFT) results for the drag and lift coefficients over a time series of 2.26 seconds at a cavitation number of 1, examining porous and non-porous hydrofoils under LES and k-ω SST turbulence models. The spectral analyses reveal that, in all cases, a dominant frequency peak is observed around 0.5 Hz. This low-frequency component represents the primary cavitation shedding frequency or cavity oscillation cycle. It is closely associated with the formation and periodic detachment of large-scale vapor cavities, a dynamic that aligns with the physical behavior of cloud cavitation. The large vortex structures linked to this frequency suggest a well-established cycle of cavity growth and collapse, confirming the periodicity of cavitation phenomena in the flow over the hydrofoil.

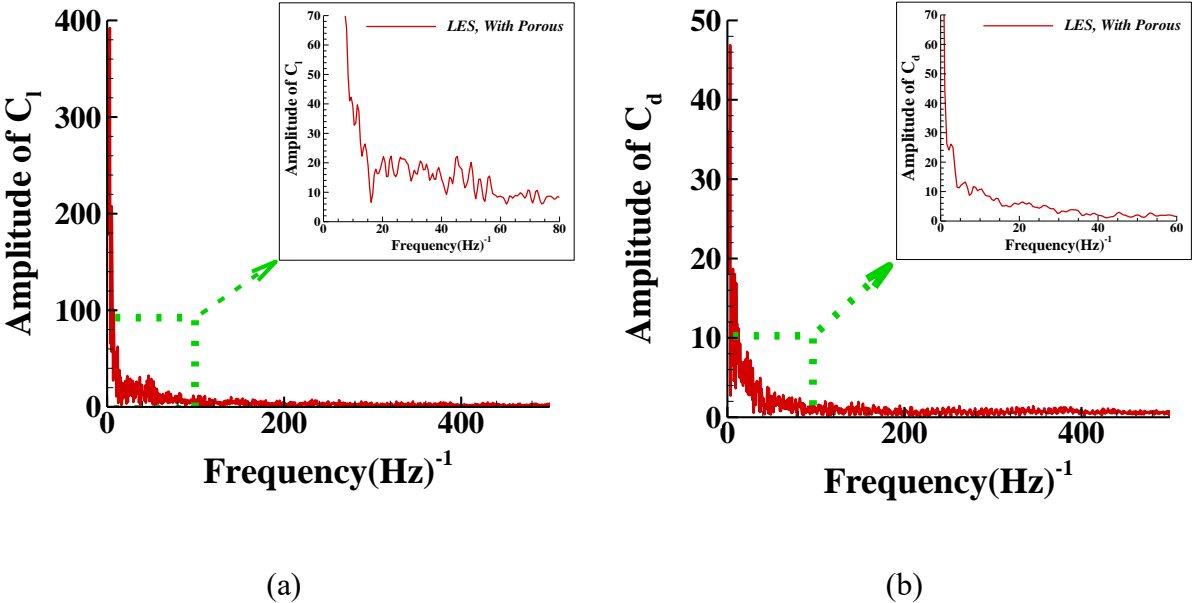

(a)          (b)

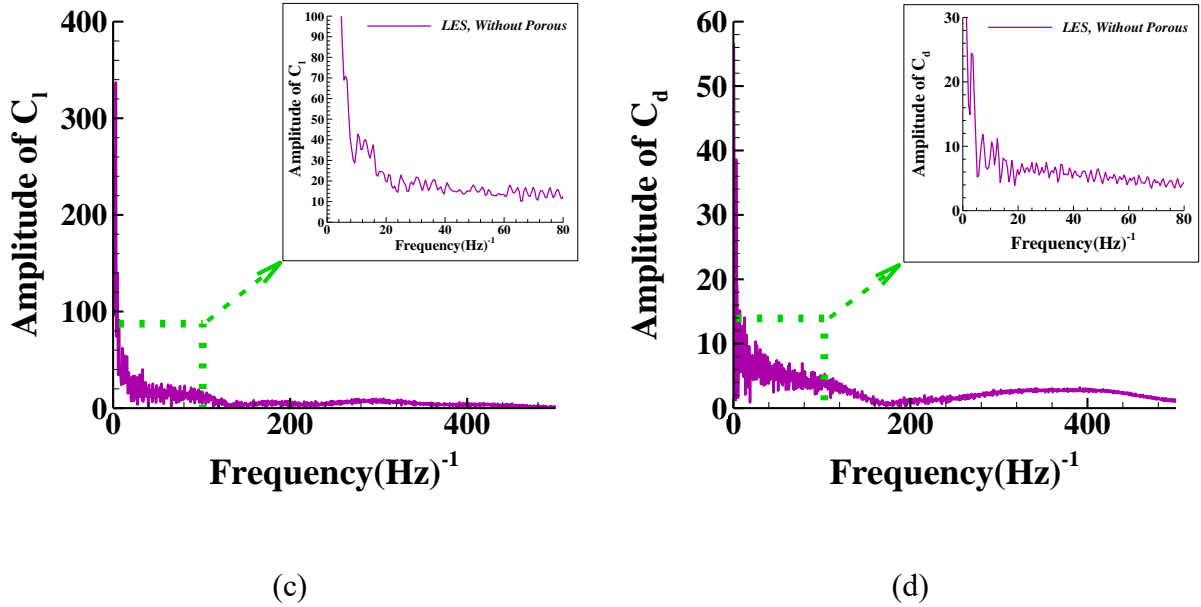

(c)                                       (d)

Fig. 13. FFT analysis of $C_l$ and $C_d$ at cavitation number 1 for (a), (b) porous and (c), (d) non-porous hydrofoil using LES turbulence model.

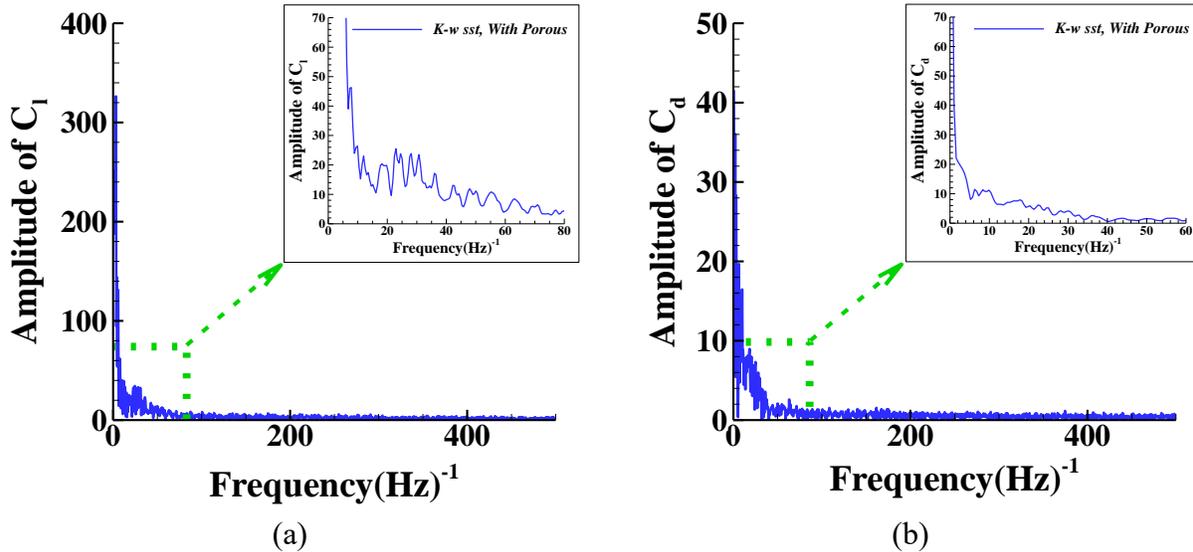

(a)                                       (b)

Fig. 14. FFT analysis of (a) $C_l$ and (b) $C_d$ at cavitation number 1 for porous hydrofoil using k-ω SST turbulence model.

In the non-porous LES case (Fig. 13c and Fig. 14b), the spectral content is significantly broader, with high amplitudes persisting up to approximately 20 Hz. This wideband spectral signature indicates a strong presence of small-scale vortical structures and more persistent vortex shedding events. The absence of an evident energy decay before 10 Hz highlights the unsteady nature of the flow. It suggests the hydrofoil surface is subject to sustained, high-frequency cavitation-induced fluctuations. These persistent harmonics imply a highly dynamic environment with frequent vapor bubble collapses and reformation cycles, ultimately leading to larger unsteady loads on the hydrofoil.

In contrast, the porous hydrofoil configurations demonstrate a markedly different spectral behavior. For both LES and k-ω SST models, a sharp decay in energy is observed beyond the dominant peak in the 0.5–5 Hz range. This indicates that the porous surface strongly influences higher-frequency content within the flow. The inherent damping properties of the porous medium serve to dampen secondary harmonics, thereby stabilizing the boundary layer and reducing the possibility of oscillations from cavitation. The energy attenuation beyond the first frequency peak significantly improves the performance of the hydrofoil in cavitating conditions.

Additionally, porous media significantly reduces variations in the $C_d$ and serves as a better measure of the unsteadiness caused by cavitation than lift. For example, in the case of a non-porous LES, the maximum amplitude of $C_d$ exceeds 50, which shows significant unsteady drag forces are created due to the relationship between the growth, collapse, and then re-growth of the vapor cavity. This larger amplitude indicates vigorous vortex shedding and considerable in-surface pressure variation across the hydrofoil surface. However, the same maximum amplitude for $C_d$ is significantly reduced with porous treatment, providing more localized amplitude distribution at lower frequencies, with lower maximum magnitude. Such reduction confirms the porous media's effectiveness in suppressing large-scale flow instabilities and reducing unsteady hydrodynamic loads. The substantial reduction in amplitude of drag fluctuations also indicates the porous layer's capacity to provide more constant boundary layer behavior and minimize the impact of pressure fluctuations due to cavitation on the surface of the hydrofoil.

Interestingly, the LES solutions of the porous hydrofoil still possess a greater extent of high-frequency content than the k-ω SST case. This is to be expected, as LES by definition resolves the smaller turbulent structures and provides a more complete capture of unsteady flow features. However, the magnitude of these high-frequency modes is considerably lower in the porous LES case compared to the non-porous one, indicating the efficient damping effect caused by the porous treatment. Although LES resolves more detailed turbulence, the porous layer successfully dampens the higher-mode energetic contribution.

Overall, porous surface treatment of the hydrofoil upper surface is very effective at controlling cavitation, damping structural vibration, and attenuating sound. Cavitation-induced instabilities and unsteady loads are reduced by porous treatment, which dampens high-frequency harmonics and stabilizes flow separation. The outcomes confirm using porous coatings or inserts as an efficient passive flow control technique to optimize hydrofoil performance in cavitating flows.

**4.4 Flow field**

Fig. 15 through Fig. 20 present contour visualizations of vapor volume fraction, velocity magnitude, and pressure coefficient for a hydrofoil treated with and without porous media, evaluated using three turbulence models, k-ε, k-ω SST, and LES, at cavitation numbers $\sigma = 1$ and $\sigma = 0.7$. These figures provide crucial insight into the influence of the porous layer on the flow structure and cavitation dynamics.

**4.4.1 Vapor volume fraction contour analysis**

The vapor volume fraction contours (Fig. 15 and Fig. 18) directly indicate cavitation bubble formation, spatial distribution, and detachment behavior. At $\sigma = 1$, porous treatment limits cavity development in all models. In particular, the k-ω SST model shows a noticeably shorter and thinner cavity profile than the k-ε model, demonstrating the former's superior response to near-wall flow stabilization induced by the porous layer. This effect is even more pronounced at $\sigma = 0.7$, where porous k-ω SST maintains a relatively confined cavitation zone. At the same time, the non-porous counterpart exhibits a thick, attached vapor sheet along the suction surface.

Fig. 15 and Fig. 18 suggest that this first observed feature, the reduction in cavity length and thickness, indicates a decrease in cavitation occurrence at higher cavitation numbers, which aligns with the findings of Scripkin et al. [104]. The porous media appears to suppress the conditions that typically trigger sheet cavitation, such as sustained low-pressure regions and boundary layer separation.

In the LES model, porous treatment significantly breaks up the coherent vapor structures into smaller, detached pockets, suggesting disruption of vapor growth and reduced energy transfer through the cavity. In contrast, the non-porous LES simulations display long, continuous cavities stretching toward the trailing edge and persisting downstream, indicating stronger cavitation collapse events. Notably, LES reduces cavity shedding frequency and forms more fragmented and damped vapor structures in the porous case, reflecting superior control over unsteady flow separation.

A critical observation is the emergence of secondary cavities and detached vapor clusters downstream of the main cavity in non-porous LES cases, especially at $\sigma = 0.7$. These features indicate secondary collapse zones and pressure rebound events, mostly absent in porous-treated configurations. This further highlights the porous media's capacity to suppress both primary cavitation and subsequent instabilities. Moreover, the porous LES cases exhibit a narrower, more controlled wake, unlike the strong vortex–cavity interactions in the non-porous case, which produce elongated and unstable vapor tails.

Furthermore, LES can resolve small-scale vortex structures and capture the cavity evolution cycle more accurately than RANS-based models. These include cavity growth, collapse, re-entrant jet formation, and cloud detachment, key features often smoothed or underpredicted in k-ε and k-ω SST simulations. Therefore, LES results offer better agreement and serve as a higher-fidelity reference when evaluating cavitation suppression strategies such as porous media applications.

**4.4.2 Velocity magnitude contour analysis**

Velocity magnitude contours (Fig. 17 and Fig. 20) support these observations by highlighting the influence of porous media on boundary layer development and flow acceleration/deceleration. In porous cases, the suction-side velocity profile is more uniform and attached, with thinner low-speed (blue) zones near the surface. These characteristics suggest a delay in flow separation and a more gradual transition to reattachment, especially in the porous k-ω SST and LES cases.

In contrast, non-porous hydrofoils generate broad low-velocity zones on the suction side and in the wake, often forming "dead zones" where reverse flow or vortex recirculation dominates. These regions arise from localized vapor accumulation and bubble collapse, which reduce local pressure and cause boundary layer thickening. This restricts liquid passage through the core flow, effectively narrowing the available flow channel, a phenomenon that acts like a nozzle. This so-called "nozzle effect" increases local velocity and corresponding pressure drops in the adjacent regions.

While such acceleration may aid in localized suction enhancement, it is typically undesirable in cavitating flows, as it intensifies pressure gradients and promotes vapor regeneration downstream. By contrast, the porous layer minimizes the dead zone's size. It mitigates abrupt acceleration, leads to more uniform velocity distribution, and improves flow attachment, ultimately enhancing hydrofoil performance by reducing unsteady lift and drag fluctuations and promoting stability in the boundary layer.

### 4.4.3 Pressure coefficient contour analysis

The pressure coefficient contours (Fig. 16 and Fig. 19) correlate strongly with the vapor field, confirming that the porous layer alleviates extreme pressure drops along the suction side. In all porous cases, especially in k-ω SST and LES models, the regions of lowest pressure are shallower and more localized than their non-porous counterparts.

In the k-ε model, applying porous treatment significantly reduces the blue low-pressure region near the leading edge and mid-chord area. It suggests that the porous layer suppresses early boundary layer detachment and reduces the extent of suction-induced vapor formation. However, in the k-ω SST non-porous case, the low-pressure region is slightly expanded compared to its porous counterpart, attributed to the model's higher sensitivity to near-wall shear production and adverse pressure gradients, which, without porous mitigation, result in a stronger suction peak. In contrast, in the LES model, the changes in negative pressure regions between porous and non-porous cases are less dramatic, as LES already captures the effect of vortex damping and bubble breakup, and the porous layer mainly refines those features rather than eliminating them.

An additional observation in Fig. 19 is the sudden increase in pressure on the suction side at $\sigma = 0.7$ in non-porous cases. These localized high-pressure (red) regions appear downstream of the cavity and are particularly evident in the non-porous k-ε and LES cases. This abrupt pressure rise corresponds to the collapse of large vapor structures, where no further cavity is present to absorb energy or buffer pressure change. In these cases, the absence of a shedding mechanism leads to an impulsive pressure rebound. Such spikes can contribute to strong structural vibrations and increased noise generation. Porous treatment dampens this rebound by interrupting cavity continuity, allowing for a more gradual pressure recovery and preventing sharp spikes.

Consequently, LES remains the most accurate among the turbulence models for capturing fine-scale vortex dynamics and transient cavity behaviors due to its resolution of sub-grid turbulent structures. As a result, the vapor cloud breakup, reattachment, and downstream wake development are more faithfully resolved in LES than in the RANS models. However, the porous treatment shows clear benefits across all turbulence models by uniformly reducing the cavity size, lowering vapor volume fraction, stabilizing velocity gradients, and softening pressure gradients.

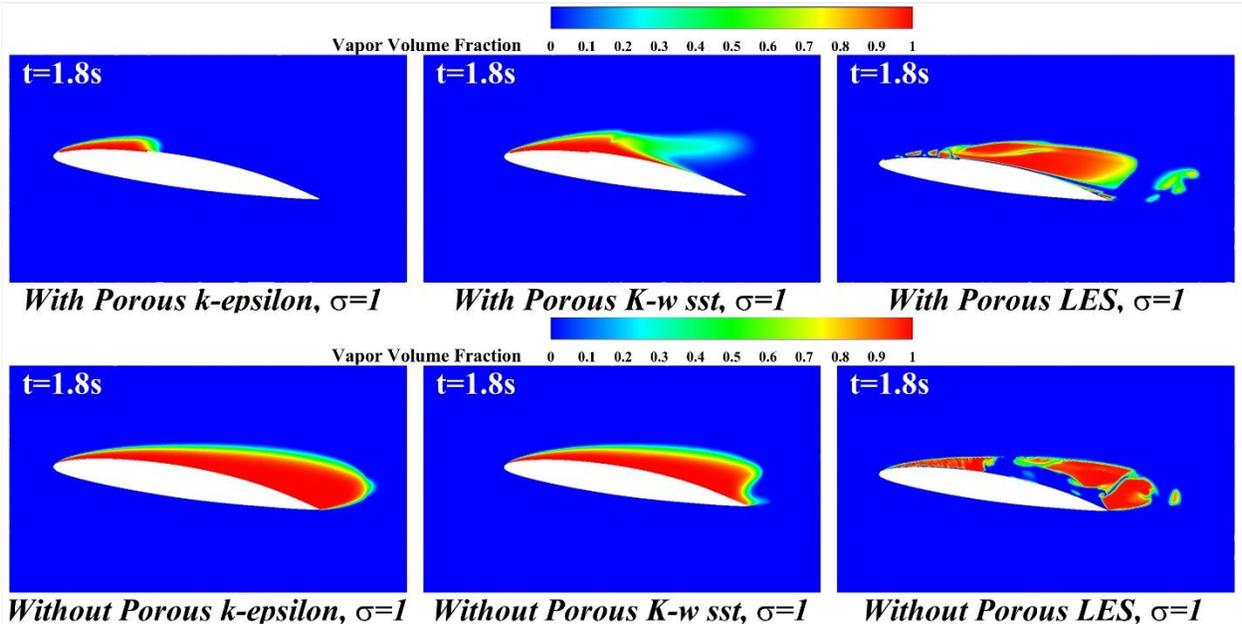

Fig. 15. Comparison of vapor volume fraction in porous and non-porous hydrofoil for cavitation number of 1 using various turbulence models.

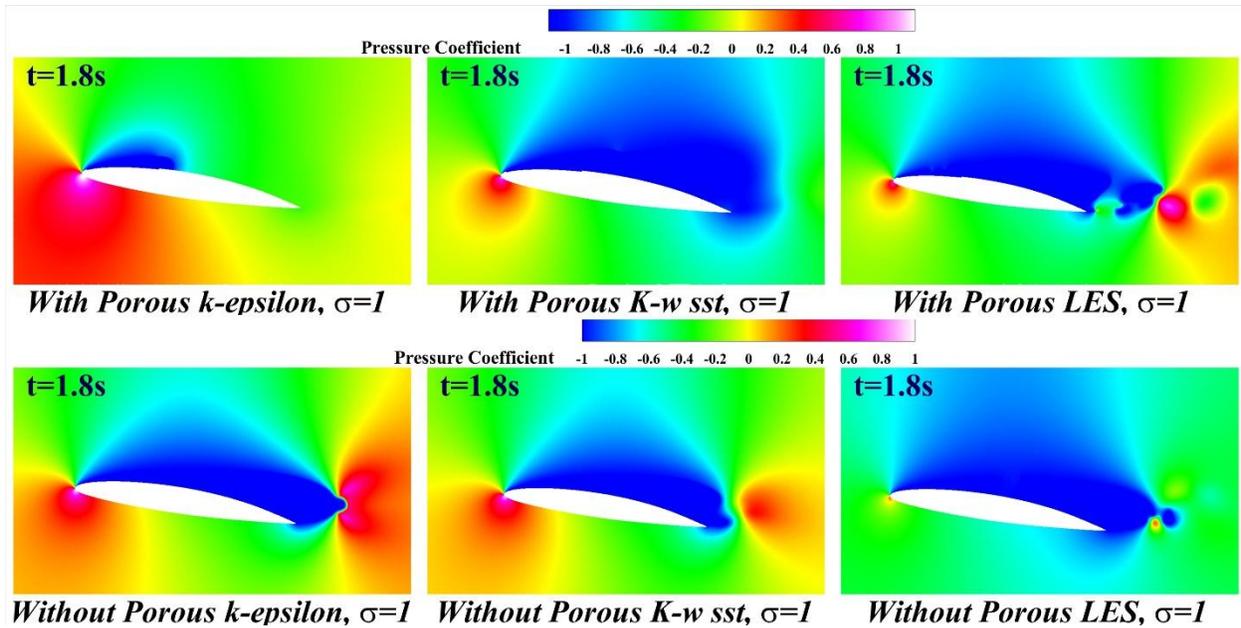

Fig. 16. Comparison of pressure coefficient contours in porous and non-porous hydrofoil for cavitation number of 1 using various turbulence models.

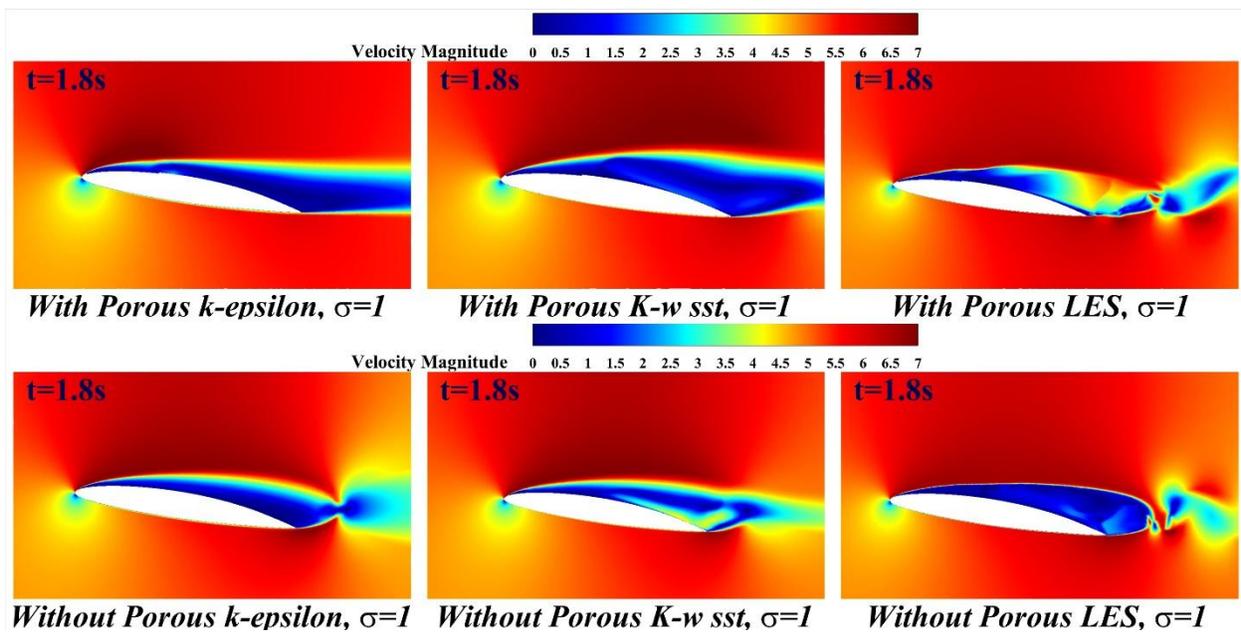

Fig. 17. Comparison of velocity magnitude in porous and non-porous hydrofoil for cavitation number of 1 using various turbulence models.

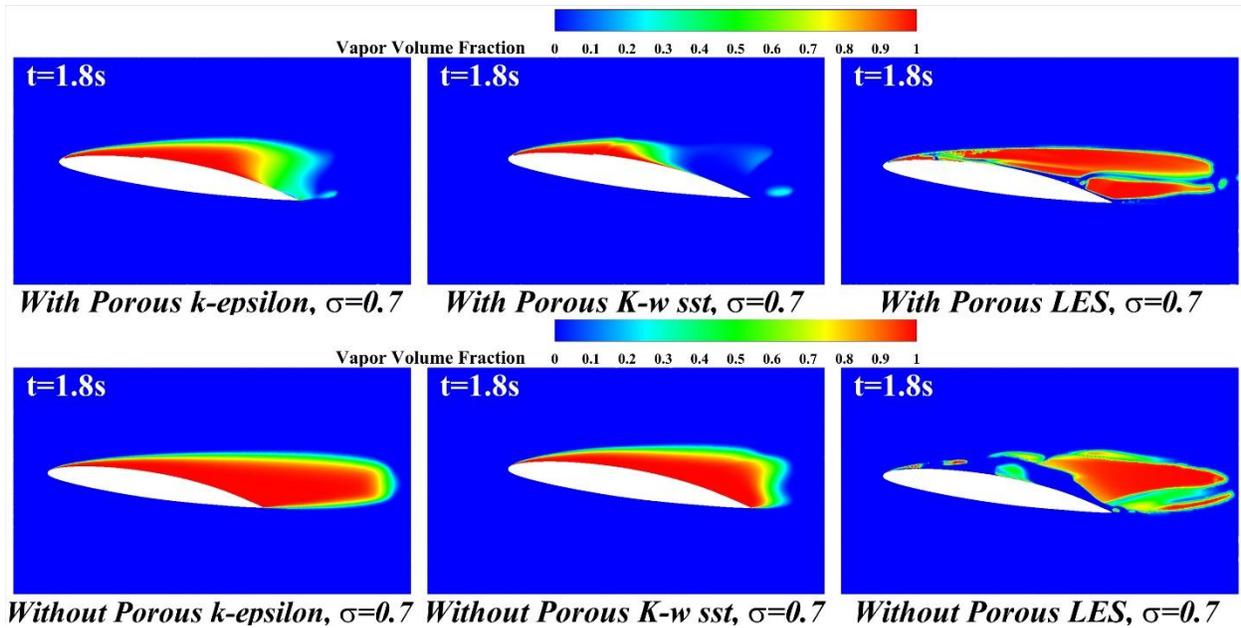

Fig. 18. Comparison of vapor volume fraction in porous and non-porous hydrofoil for cavitation number of 0.7 using various turbulence models.

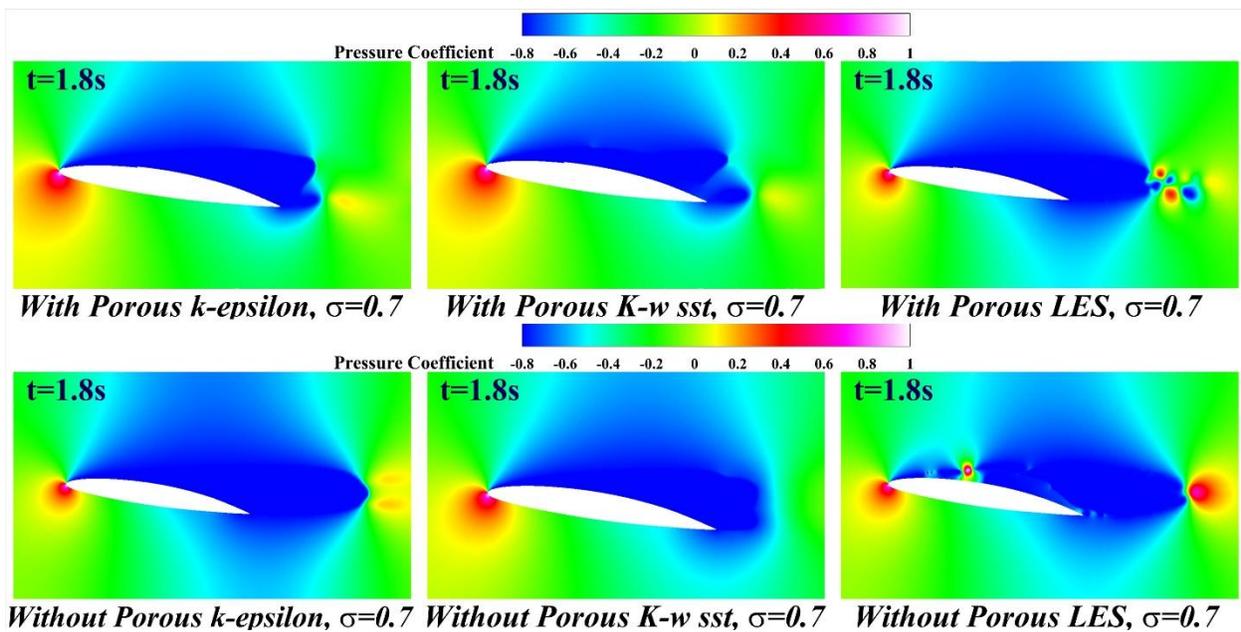

Fig. 19. Comparison of pressure coefficient contours in porous and non-porous hydrofoil for cavitation number of 0.7 using various turbulence models.

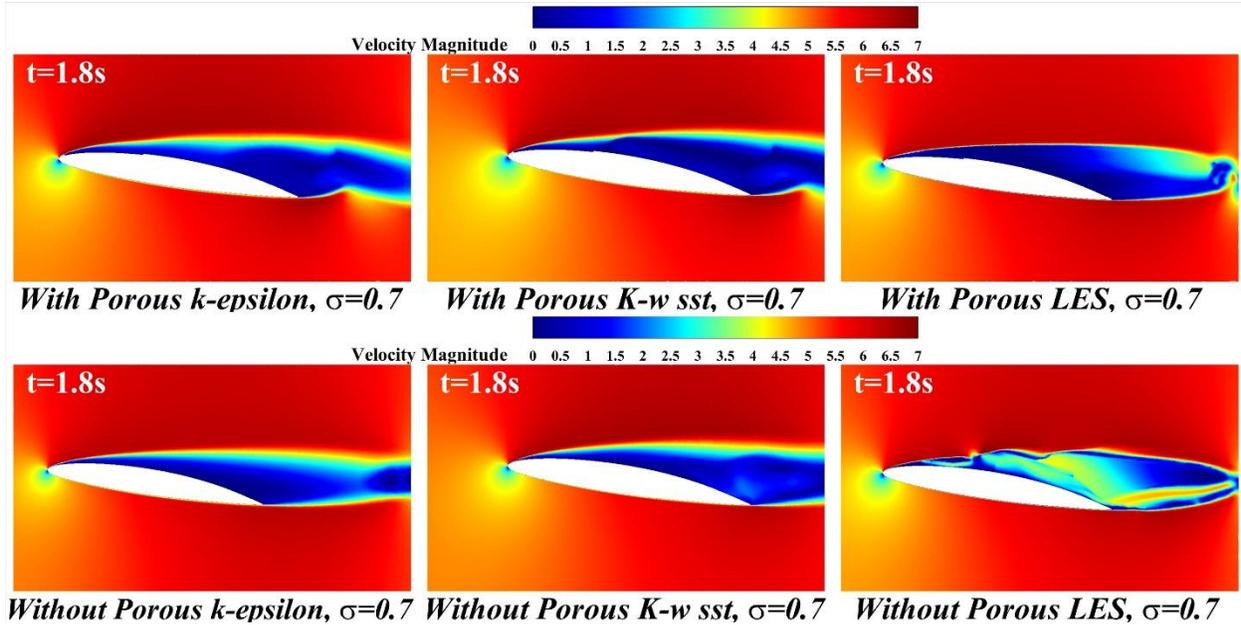

Fig. 20. Comparison of velocity magnitude in porous and non-porous hydrofoil for cavitation number of 0.7 using various turbulence models.

### 4.5 Turbulent kinetic energy (TKE) distribution analysis

Fig. 21 and Fig. 22 present the normalized turbulent kinetic energy (TKE) distributions around the hydrofoil with and without porous treatment for cavitation numbers $\sigma = 1$ and $\sigma = 0.7$, respectively. These contours show details on turbulence production, shear layer developments, and the influence of porous treatment on flow instabilities during cavitation. In cavitating flows, shear stress is critical for initiating the necessary turbulence, because near boundary layers transitioning from laminar to turbulent flows occur under adverse pressure gradients. The cavity interface is critical for transition due to unsteady phase change, recirculation, and reattachment processes. As cavitation bubbles collapse or shed downstream, turbulent structures are initiated through shear between reverse flows and the main stream, amplifying energy transfer and unsteadiness.

TKE quantifies the intensity of turbulence within a fluid flow. It represents the average kinetic energy per unit mass associated with the eddies in a turbulent regime. In essence, TKE captures the energy contained in the velocity fluctuations of fluid particles and is commonly calculated using the root mean square of these fluctuations. In fluid dynamics, particularly in the study of turbulent flows, TKE provides a key measure of the fluctuating component of motion. The following equation gives its standard mathematical formulation:

$$\text{Kinetic energy} = \frac{1}{2}mv^2 \tag{49}$$

Here, $v$ and $m$ represent velocity and mass, respectively. In turbulent flows, where the velocity field is typically described by its time-averaged components $u'$, $v'$, and $w'$, the turbulent kinetic energy (k) is defined as:

$$k = (\langle u'^2 \rangle + \langle v'^2 \rangle + \langle w'^2 \rangle)/2 \tag{50}$$

Numerous studies have explored TKE behavior in cavitating flows around hydrofoils [105]. These investigations underscore the critical role of cavitation-vortex interactions in TKE transport and the influence of cavitation on how turbulent energy is distributed across different length and time scales [106].

Across all models, elevated TKE zones appear downstream of the cavity closure region, where vapor collapse interacts with shear layers. In porous configurations, particularly with k-ε and k-ω SST models, the regions behind the hydrofoil exhibit notably lower TKE at $\sigma = 1$ (see Fig. 21). This suggests that the porous media dampen shear layer amplification, delay flow separation, and reduce vortex strength, thereby suppressing wake instability. Consequently, fewer or no coherent vortices are formed in these downstream regions for porous RANS models under higher cavitation numbers.

However, this stabilizing behavior weakens at $\sigma = 0.7$, where more intense cavitation leads to longer cavity extensions and sharper collapses. In this case (Fig. 22), even porous k-ε and k-ω SST models exhibit vortex structures in the regions behind the hydrofoil. This behavior arises because:

- Cavitation-induced pressure gradients become stronger, overwhelming some of the damping effects of the porous medium.

- Under such aggressive cavitation, RANS models, which rely on averaged turbulence quantities, tend to produce abrupt cavity closure and stronger downstream shear layers.

- This shear roll-up results in visible vortex structures, even with porous walls.

At the same time, the collapse of cavity bubbles reinforces the boundary layer and reduces the likelihood of flow separation. This mechanism is especially evident at lower cavitation numbers, where intensified cavity activity contributes to localized pressure rise and promotes reattachment, thereby reducing turbulence generation. This dynamic interplay between cavity behavior, shear stress, and turbulence suppression reflects the complex coupling in cavitating flows [107].

In contrast, the LES model with porous treatment continues to show a smoother and more diffused TKE field, even under low cavitation number conditions. LES's ability to resolve sub-grid turbulent eddies leads to gradual vapor interface breakup, less abrupt reattachment, and thus reduced coherence of downstream shear structures. Consequently, the porous LES model prevents distinct vortex formation in the region behind the hydrofoil, preserving stability better than its RANS counterparts.

A critical observation is the emergence of secondary cavities and detached vapor clusters downstream of the main cavity in non-porous LES cases, especially at $\sigma = 0.7$. These features indicate secondary collapse events and pressure rebound zones, which are largely absent in porous-treated models. This highlights the porous media's ability to suppress primary cavitation and subsequent flow instabilities.

Furthermore, the porous LES configurations display a narrower and more stable wake, while non-porous LES generates extended vortex–cavity interactions, visible as long turbulent trails in the downstream region.

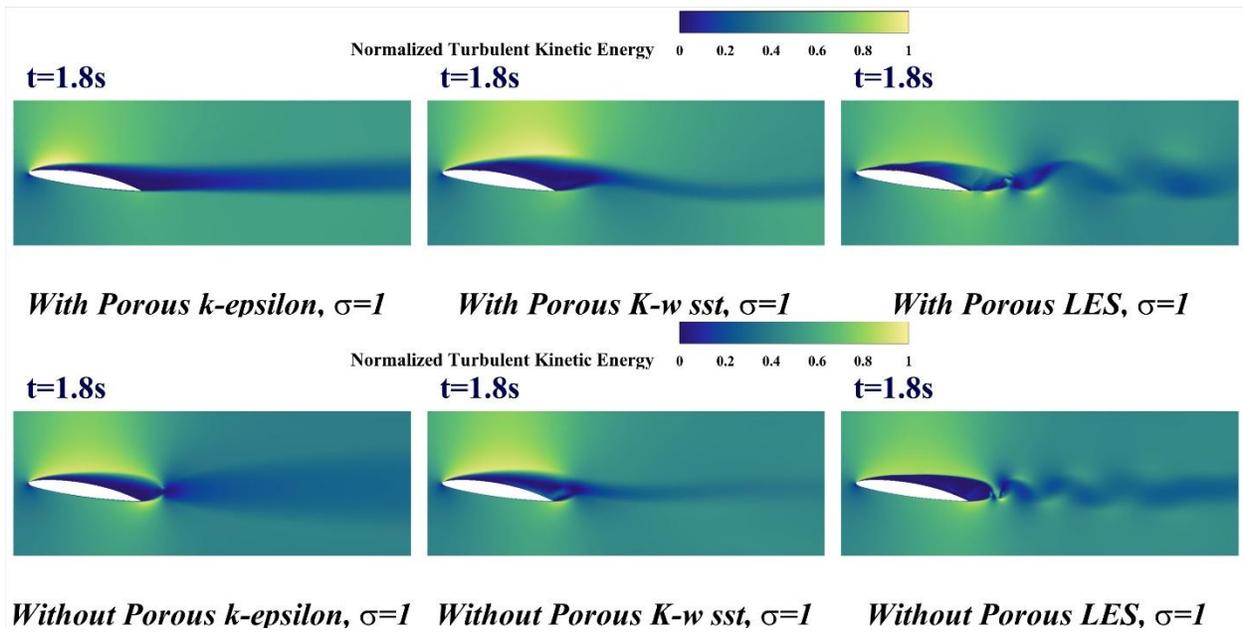

Fig. 21. Comparison of normalized turbulent kinetic energy contours in porous and non-porous hydrofoil for cavitation number of 1 using various turbulence models.

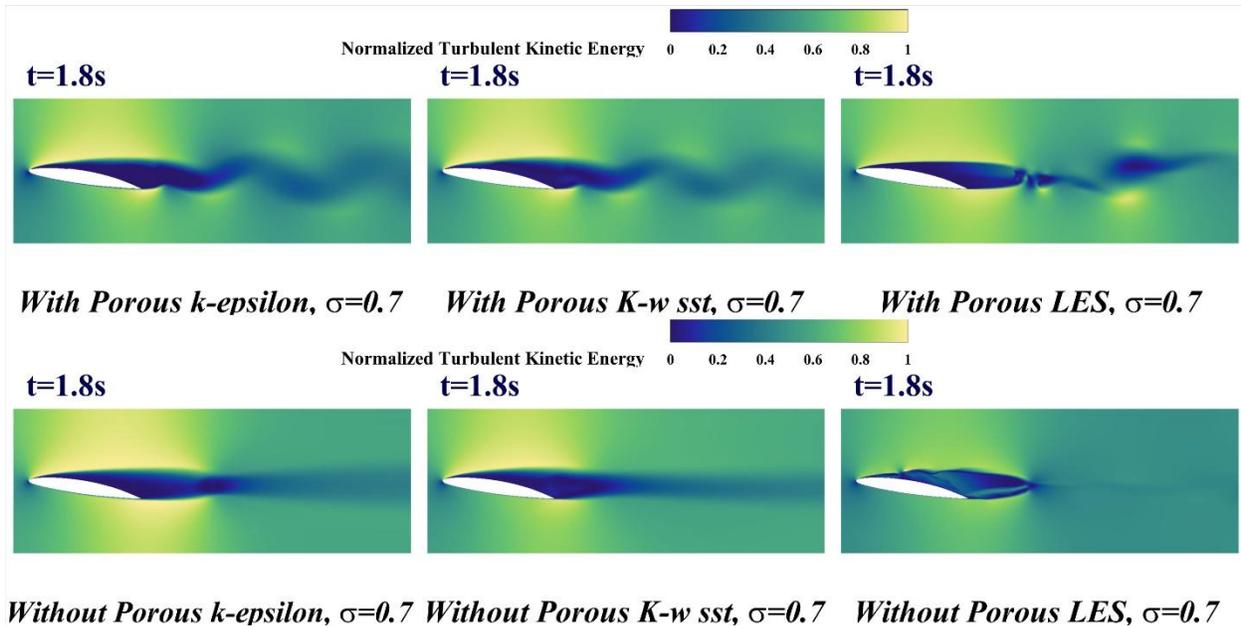

Fig. 22. Comparison of normalized turbulent kinetic energy contours in porous and non-porous hydrofoil for cavitation number of 0.7 using various turbulence models.

### 4.6. Vorticity transport equation

The vorticity transport equation provides critical insights into the behavior of cavitating flows and is instrumental in analyzing various fluid dynamics scenarios. In viscous flows subjected to a no-slip boundary condition, such as those around a hydrofoil, velocity distributions reveal the significance of vorticity stretching and dilatation in vortex formation. Understanding these mechanisms is essential for optimizing pump performance and enhancing durability [108].

Research on twisted hydrofoils has shown that shear layer separation and re-entrant jets drive multiscale vortex dynamics through mechanisms like stretching and dilatation [109]. Additionally, the porous hydrofoil generates varying cavitation regimes at different cavitation numbers, influencing the collapse dynamics of cavitation clouds. These collapsing structures subsequently generate eddies throughout the flow field.

With a zero-velocity boundary at the wall, rotational motion is induced within the adjacent boundary layer [109]. Considering the incompressible and viscous nature of the flow, Helmholtz's vorticity theory can be applied to model the rotational behavior of the fluid. This approach enables the analysis of vortex-cavitation interactions near a porous hydrofoil, focusing on rotational motion in the direction normal to the two-dimensional Cartesian plane. The governing vorticity transport equation is expressed as:

$$\frac{D\omega_z}{Dt} = [(\vec{\omega}\cdot\nabla)\vec{V}]_z - [(\vec{\omega}(\nabla\cdot\vec{V}))]_z + [\frac{\nabla\rho_m \times \nabla p}{\rho_m^2}]_z + [(\upsilon_m + \upsilon_t)\nabla^2\vec{\omega}]_z \qquad (51)$$

$$\omega_z = \frac{\partial V_y}{\partial x} - \frac{\partial V_x}{\partial y} \qquad (52)$$

In the equation (51), the term $\frac{D\omega_z}{Dt}$ represents the rate of change of vorticity, while $[(\vec{\omega}\cdot\nabla)\vec{V}]_z$ denotes vorticity stretching due to velocity gradients. The $[(\vec{\omega}\cdot\nabla)\vec{V}]_z$ is further defined as:

$$[(\vec{\omega}\cdot\nabla)\vec{V}]_z = \omega_x \frac{\partial V_z}{\partial x} + \omega_y \frac{\partial V_z}{\partial y} + \omega_z \frac{\partial V_z}{\partial z} \qquad (53)$$

However, because this study focuses on two-dimensional flow, the stretching term is identically zero. In 2D flows, the vorticity vector lies perpendicular to the flow plane, while the velocity and its gradient are confined within the plane. This orthogonal arrangement leads to the vanishing of the vorticity stretching term.

Equation (51) also includes the vorticity dilatation term, which captures the effect of fluid compressibility on vorticity due to volumetric expansion or contraction [110]. This term is negligible in non-cavitating flows, with minimal relative velocity differences. However, in cavitating regimes, it becomes significant [111]. As shown in numerical simulations by Ji et al. [10], cavitation enhances mass transfer rates, intensifying vorticity generation and promoting rotational motion within the boundary layer.

$$[(\vec{\omega}(\nabla \cdot \vec{V})]_z = \omega_z \left( \frac{\partial V_x}{\partial x} + \frac{\partial V_y}{\partial y} + \frac{\partial V_z}{\partial z} \right) \tag{54}$$

Equation (54) incorporates the baroclinic torque term, which arises when pressure and density gradients are misaligned.

$$[\frac{\nabla \rho_m \times \nabla p}{\rho_m^2}]_z = \frac{1}{\rho_m^2} \left( \frac{\partial \rho_m}{\partial x} \cdot \frac{\partial p}{\partial y} - \frac{\partial \rho_m}{\partial y} \cdot \frac{\partial p}{\partial x} \right) \tag{55}$$

The coupling between velocity divergence and spatial variations in pressure and density, both driven by cavitation, makes vorticity dilatation and baroclinic torque key contributors to vortex formation. These terms are crucial for understanding the evolution of the vorticity field in unstable cavitating flows [111]. Although the final term in the equation (51), representing viscous diffusion, also affects vorticity; its influence is typically much weaker than that of the stretching, dilatation, and baroclinic contributions.

**4.6.1 Vorticity generation and baroclinic torque analysis**

Fig. 23 through Fig. 25 present the contours of baroclinic torque, vorticity dilatation, and z-vorticity for porous and non-porous hydrofoils under various turbulence models. These quantities are critical in describing cavitating flows' rotational characteristics and unsteady dynamics.

In the Z-vorticity contours, blue regions represent negative (clockwise) vortex rotation, while red regions indicate positive (counterclockwise) rotation. These patterns arise due to flow separation, cavitation bubble collapse, and reattachment processes on the suction side of the hydrofoil. As evident in the non-porous LES case, strong alternating vortex structures are shed from the trailing edge, creating a highly unstable wake. In contrast, porous-treated configurations, especially under k-ε and k-ω SST, show significantly weakened or absent vortex structures in the same downstream regions. This confirms the effectiveness of porous surfaces in attenuating wake instabilities and reducing vortex strength, a key contributor to minimizing flow-induced vibrations and noise. Interestingly, in cavitation number $\sigma = 0.7$, some porous k-ε and k-ω SST cases still show wake vortices, unlike LES, indicating that LES suppresses vortex regeneration more effectively under severe cavitation conditions.

The baroclinic torque contours illustrate rotational forces induced by the misalignment of pressure and density gradients during cavity collapse. Jahanbakhshi et al. [112] noted that this torque contributes to rotational motion around collapsing cavities. In our figures, non-porous LES and k-ω SST cases demonstrate large baroclinic torque values, especially near the cavity closure and reattachment regions, where abrupt density changes occur. In contrast, porous configurations show substantially reduced baroclinic torque magnitudes, indicating the suppression of rotational energy sources. While baroclinic torque influences vortex formation, its effect is often secondary to mechanisms like vorticity dilatation and vortex stretching, consistent with findings by Yang et al. [113]. These observations align with Laberteaux and Ceccio's findings [114], who also emphasized baroclinic torque's limited but non-negligible role relative to other vorticity generation mechanisms in cavitating flows.

The vorticity dilatation contours further emphasize the complex dynamics inside the cavity. Blue regions denote negative dilatation, corresponding to compression and vortex contraction during cavity collapse. Red regions highlight positive dilatation, arising from cavity expansion and flow divergence. These dynamics play a primary role in vorticity generation within cavitating flows. The porous hydrofoils reduce the magnitude and spatial extent of both positive and negative dilatation zones, stabilizing the boundary layer and mitigating flow separation. This effect is significant in LES results, where fine-scale dilatation features are sharply resolved, especially in the non-porous case. While still capturing transient structures, the porous LES exhibits more localized and moderate dilatation fields, supporting the idea of damped cavity-induced unsteadiness.

These observations are consistent with Schnerr et al. [7], who successfully simulated cavitation collapse using a barotropic model and highlighted the importance of vorticity dilatation over other rotational sources. In the present study, it becomes evident that porous treatment helps reduce the spatial range and magnitude of these rotational mechanisms. The relative vorticity dilatation, closely related to the velocity divergence field, transports vorticity from the outer flow into the cavity and reshapes the flow's rotational topology. This process is visibly more contained in the porous cases, further reinforcing the hydrofoil's dynamic stability.

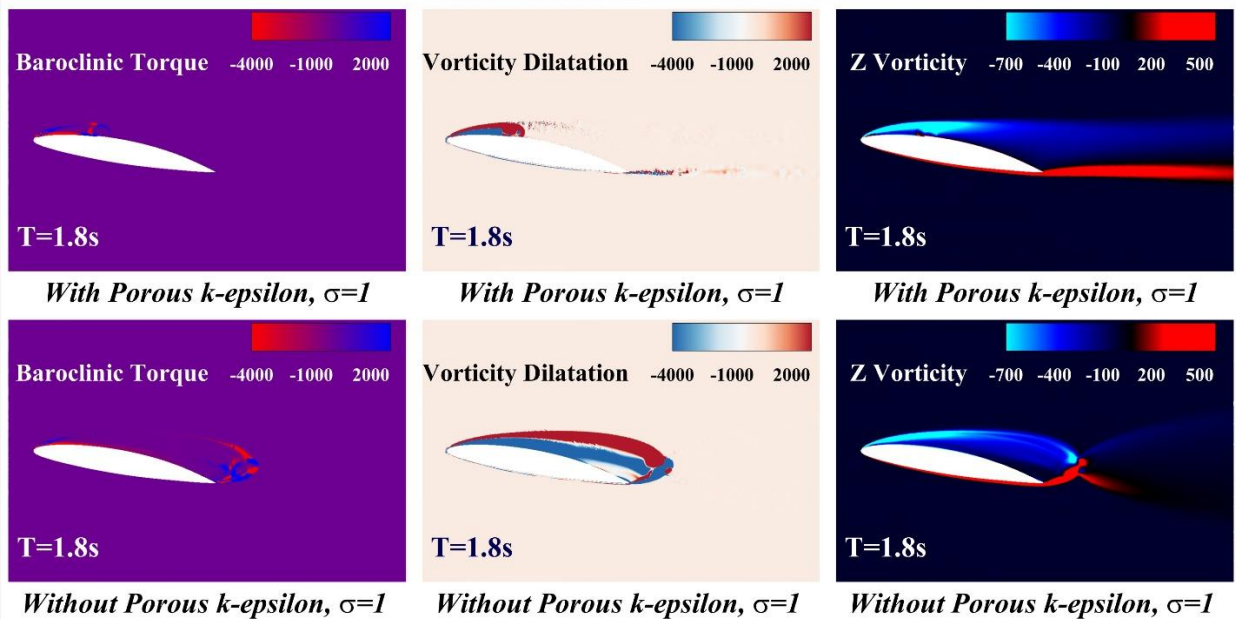

Fig. 23. Analysis of cavitation effects on vorticity contours normal to the 2D plane, baroclinic torque, and vorticity dilatation for porous and non-porous hydrofoils using the k-ε turbulence model at a cavitation number of 1.

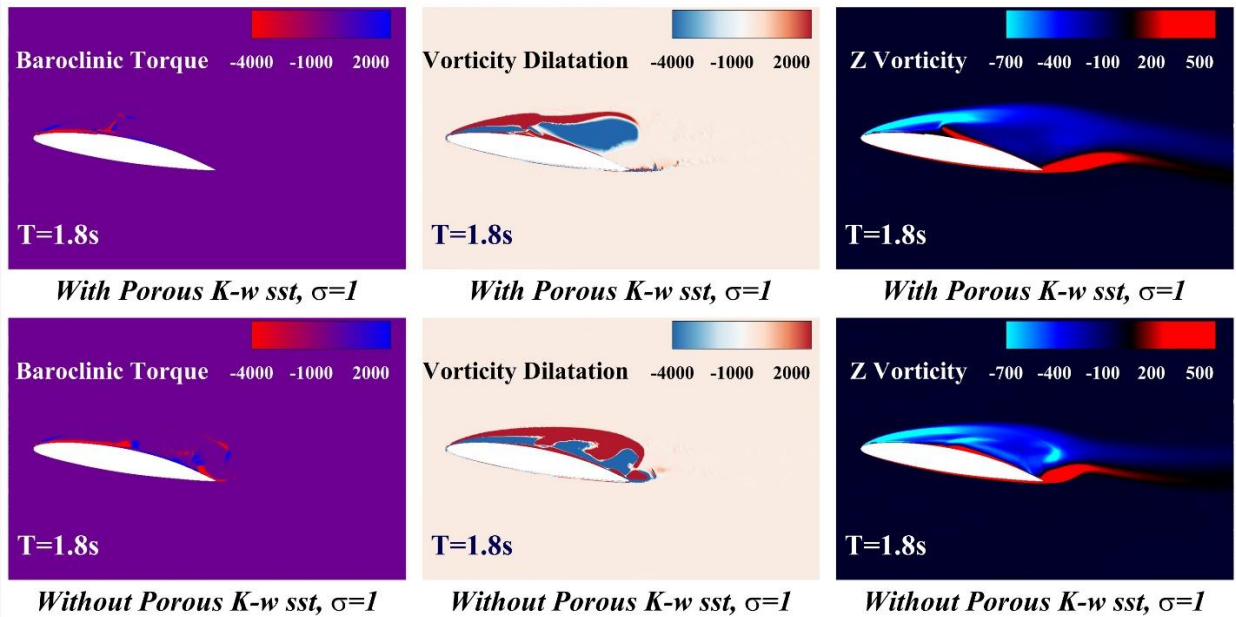

Fig. 24. Analysis of cavitation effects on vorticity contours normal to the 2D plane, baroclinic torque, and vorticity dilatation for porous and non-porous hydrofoils using the k-ω SST turbulence model at a cavitation number of 1.

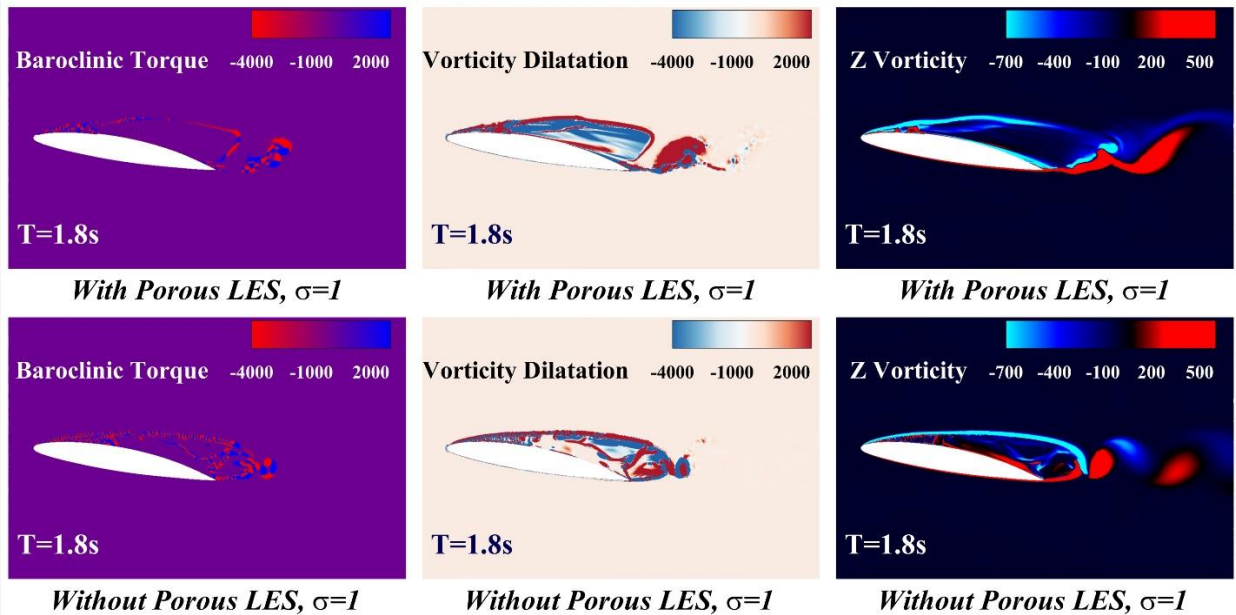

Fig. 25. Analysis of cavitation effects on vorticity contours normal to the 2D plane, baroclinic torque, and vorticity dilatation for porous and non-porous hydrofoils using the LES turbulence model at a cavitation number 1.

## 5. Conclusions

This study employs Computational Fluid Dynamics (CFD) to investigate passive cavitation control using porous surface treatments on a NACA 66 (MOD) hydrofoil. The simulations combine the Schnerr–Sauer cavitation model with a VOF interface tracking method to capture phase interactions accurately. For 2D analyses, the study uses three turbulence models, Large Eddy Simulation (LES), realizable k-ε, and SST k-ω, to compare their predictive capabilities. In contrast, the 3D simulation focuses on spanwise cavitation dynamics, employing the LES model due to its proven accuracy in near-wall flows. The performance and effectiveness of porous surfaces were analyzed using several diagnostics, including pressure coefficient profiles, vapor volume fraction, velocity and pressure contours, FFT of hydrodynamic forces, and field-based quantities such as vorticity, baroclinic torque, and turbulent kinetic energy (TKE). Key findings are summarized below:

1. **Porous Media Enhances Cavitation Suppression Across Models and Dimensions**
   The porous hydrofoil configuration consistently demonstrates suppressed cavitation activity and enhanced flow stability across all tested models. LES results in shorter and fragmented vapor cavities and delayed shedding. k-ω SST accurately captures smoother wake reattachment, and the realizable k-ε model, despite its diffusive nature, also shows improved pressure recovery with porous treatment.

2. **Flow Stabilization and Pressure Gradient Mitigation**
   The LES-based analysis reveals that porous treatment delays cavitation shedding, promotes smoother wake development, and flattens pressure recovery zones, which are critical for reducing pressure pulses and unsteady loading. The wake becomes less turbulent and more coherent. In contrast, non-porous surfaces exhibit high-frequency pressure spikes and aggressive cloud shedding. These dynamics confirm that the porous layer diffuses energy associated with re-entrant jets and cavity collapses.

3. **Impact on Pressure and Velocity Fields**

   Contour plots of velocity magnitude and pressure coefficient reveal that porous media prevent abrupt acceleration and pressure drops near the hydrofoil suction side. In non-porous cases, the porous-treated surface sustains a stable vapor layer near the wall with milder suction. Velocity profiles show a reduction in low-speed recirculation zones and delayed flow separation. These modifications result in more gradual and localized cavity collapse, decreasing the likelihood of structural vibration and noise generation.

4. **Vortex Attenuation and Wake Control**

   Porous treatment reduces z-vorticity and baroclinic torque magnitude, particularly in the downstream regions. The non-porous LES case generates strong alternating vortex structures that destabilize the wake. Conversely, porous LES and RANS configurations exhibit a diffused, lower-energy wake field with diminished vortex strength. This suppression of vortex formation confirms the role of porous media in limiting rotational instabilities and enhancing flow coherence.

5. **TKE Suppression and Spectral Damping**

   FFT analysis reveals significantly reduced spectral energy beyond the primary cavitation frequency (around 0.5 Hz) in porous cases. The porous LES case exhibits lower amplitude harmonics across various frequencies, indicating reduced high-frequency content. TKE fields further validate these findings: porous surfaces lower turbulent kinetic energy behind the hydrofoil, especially at $\sigma = 1$. Although cavitation intensifies at $\sigma = 0.7$, the porous LES setup suppresses vortex reformation more effectively than its RANS counterparts.

**Appendix A**

We assess the sensitivity of the Smagorinsky coefficient ($C_s$) on the accuracy of our simulations. A comparable approach was adopted by Bin Ji et al. [115], whose work closely aligns with the present study. Accordingly, our baseline simulations selected the same numerical value for $C_s$ to ensure consistency and comparability. To evaluate its influence, we conducted additional simulations using the coefficient's lower and upper bound values. The study retained the initially selected value, which demonstrated stable performance throughout. Fig. 26 illustrates a comparative summary of these tests.

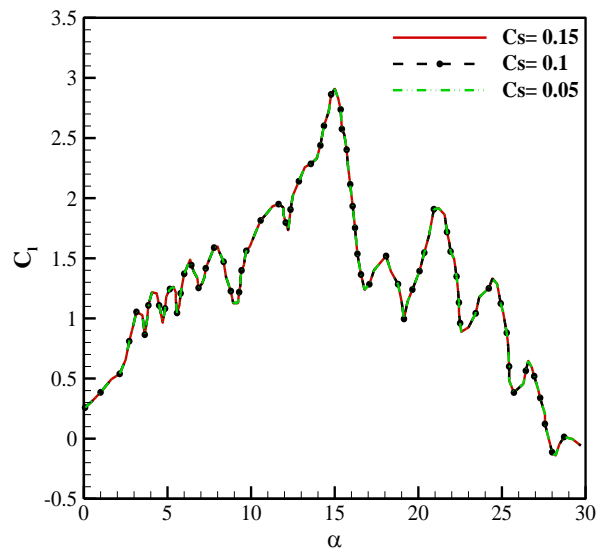

Fig. 26. $C_s$ sensitivity investigation for 2D LES simulation